\documentclass[%
nofootinbib,
 amsmath,2amssymb,
prd,
twocolumn,
10pt
]{revtex4-1}
\usepackage[bottom]{footmisc}
\usepackage[utf8]{inputenc}
\usepackage{slashed}
\usepackage[%dvipdfm,
colorlinks=true,
linkcolor=blue,
breaklinks=true,
urlcolor=blue,
citecolor=blue]{hyperref}
\usepackage{multirow}
\usepackage{diagbox}
\usepackage{dcolumn}% Align table columns on decimal point
\usepackage{bm}% bold math
\usepackage{hyperref}% add hypertext capabilities
\usepackage{setspace}
\usepackage{makecell}
\usepackage{threeparttable}
\usepackage{footnote}
\usepackage{graphicx}% Include figure files
\usepackage{float}
\allowdisplaybreaks[4]
\newcommand{\bra}[1]{\left\langle #1 \right|}
\newcommand{\ket}[1]{\left| #1 \right\rangle}

\newcommand{\lra}[1]{\left(#1\right)}
\newcommand{\lrb}[1]{\left[#1\right]}

\usepackage{relsize}
\def\babar{\mbox{\slshape B\kern-0.1em{\smaller A}\kern-0.1em
    B\kern-0.1em{\smaller A\kern-0.2em R}}}

\begin{document}

\title{A survey of heavy-antiheavy hadronic molecules}

\author{Xiang-Kun Dong$^{1,2}$}\email{dongxiangkun@itp.ac.cn}
 \author{Feng-Kun Guo$^{1,2}$} \email{fkguo@itp.ac.cn}
\author{Bing-Song Zou$^{1,2,3}$}\email{zoubs@itp.ac.cn}

\address{%
$^1$CAS Key Laboratory of Theoretical Physics, Institute of Theoretical Physics,\\
Chinese Academy of Sciences, Beijing 100190, China\\
$^2$School of Physical Sciences, University of Chinese Academy of Sciences, Beijing 100049, China\\
$^3$School of Physics, Central South University, Changsha 410083, China
}

\begin{abstract}
Many efforts have been made to reveal the nature of the overabundant resonant structures observed by the worldwide experiments in the last two decades. Hadronic molecules attract special attention because many of these seemingly unconventional resonances are located close to the threshold of a pair of hadrons. To give an overall feature of the spectrum of hadronic molecules composed of a pair of heavy-antiheavy hadrons, namely, which pairs are possible to form molecular states, we take charmed hadrons for example to investigate the interaction between them and search for poles by solving the Bethe-Salpeter equation. 
We consider all possible combinations of hadron pairs of the $S$-wave singly-charmed mesons and baryons as well as the narrow $P$-wave charmed mesons.
The interactions, which are assumed to be meson-exchange saturated, are described by constant contact terms which are resummed to generate poles. It turns out that if a system is attractive near threshold by the light meson exchange, there is a pole close to threshold corresponding to a bound state or a virtual state, depending on the strength of interaction and the cutoff. In total, 229 molecular states are predicted. The observed near-threshold structures with hidden-charm, like the famous $X(3872)$ and $P_c$ states, fit into the  spectrum we obtain. We also highlight a $\Lambda_c\bar \Lambda_c$ bound state that has a pole consistent with the cross section of the $e^+e^-\to\Lambda_c\bar \Lambda_c$ precisely measured by the BESIII Collaboration.

\end{abstract}

\maketitle

\medskip

\tableofcontents

\medskip

\section{Introduction}

The formation of hadrons from quarks and gluons is governed by quantum chromodynamics (QCD), which at low energy is nonperturbative. Therefore, it is difficult to tackle with the hadron spectrum model-independently. The traditional quark model~\cite{GellMann:1964nj,Zweig:1964jf}, where hadrons are classified as mesons, composed of $q\bar q$ and baryons, composed of $qqq$, provides quite satisfactory description of the hadrons observed in last century. The last two decades witnessed the emergence of plenty of states or resonant structures in experiments, many of which do not fit the hadron spectrum predicted by the naive quark model and thus are candidates of the so-called exotic states. Many efforts have been devoted to understand the nature of these states but  most of them still remain controversial (see Refs.~\cite{Chen:2016qju,Hosaka:2016pey,Richard:2016eis,Lebed:2016hpi,Esposito:2016noz,Guo:2017jvc,Ali:2017jda,Olsen:2017bmm,Kou:2018nap,Kalashnikova:2018vkv,Cerri:2018ypt,Liu:2019zoy,Brambilla:2019esw,Guo:2019twa,Yang:2020atz,Ortega:2020tng} for recent reviews). 

The observation that most of these structures are located near the threshold of a pair of heavy-antiheavy hadrons may shed light on identifying the nature of these structures.
To name a few famous examples, let us mention the $X(3872)$, aka $\chi_{c1}(3872)$~\cite{Choi:2003ue} and $Z_c(3900)^\pm$~\cite{Ablikim:2013mio,Liu:2013dau,Ablikim:2013xfr} around the $D\bar D^*$ threshold, the $Z_c(4020)^\pm$~\cite{Ablikim:2013emm,Ablikim:2013wzq} near the $D^*\bar D^*$ threshold, the $Z_b(10610)^\pm$ and $Z_b(10650)^\pm$~\cite{Belle:2011aa,Garmash:2015rfd} near the $B\bar B^*$ and $B^*\bar B^*$ thresholds, the $Z_{cs}(3985)^-$~\cite{Ablikim:2020hsk} near the $\bar D_s D^*$ and $\bar D_s^* D$ thresholds, and the $P_c$ states~\cite{Aaij:2019vzc} near the $\bar D^{(*)}\Sigma_c$ thresholds, respectively. These resonant structures were widely investigated in many models, including assigning them to be the molecular states of the corresponding systems. We refer to Ref.~\cite{Guo:2017jvc} for a comprehensive review of hadronic molecules.

Although these near-threshold resonant structures have been explored by many works using various methods, our understanding of these structures will greatly benefit from a whole and systematic spectrum of heavy-antiheavy hadronic molecules based on one single model. In this paper, we provide such a spectrum of hadronic molecules composed of a pair of heavy-antiheavy hadrons including $S,P$-wave heavy mesons and $S$-wave heavy baryons.  The interactions between these hadron pairs are assumed to be contact terms saturated by meson exchanges as in, e.g., Refs.~\cite{Ecker:1988te,Epelbaum:2001fm,Peng:2020xrf}. 
In order to have a unified treatment in all systems, we will not consider coupled-channel effects.
In Ref.~\cite{Peng:2020xrf} a similar study, focusing on the possible bound states composed of $\bar D^{(*)}$ and $\Sigma_c^{(*)}$, are performed. Also note that such interactions have been obtained in many other works, e.g. Refs.~\cite{Liu:2007bf,Ding:2009vd,Wu:2010jy,Wu:2010vk,Yang:2011wz,Valderrama:2019sid}, to look for possible bound states associated with the near-threshold resonant structures. These works used different models and conventions and in some cases the results are inconsistent with each other.

In this work we consider the resonance saturation of the contact interaction due to the one-boson-exchange (light pseudoscalar and vector mesons) together with heavy quark spin symmetry (HQSS), chiral symmetry and SU(3) flavor symmetry, to estimate the potentials of the heavy-antiheavy systems, and special attention is paid to the signs of coupling constants to get self-consistent results.  Heavy quark flavor symmetry (HQFS) implies that the potentials between bottomed hadron pairs are the same as those between charmed ones and thus we take charmed hadrons for example.  The obtained potentials are used to solve the Bethe-Salpeter equation in the on-shell form to determine the pole positions of these heavy-antiheavy hadron systems and a systematic spectrum of these hadronic molecules is obtained.

Besides the possible molecular states, the interaction between a pair of heavy-antiheavy hadrons at threshold is crucial to understand the line shape of the invariant mass distribution of related channels, as discussed in Ref.~\cite{Dong:2020hxe}. It is known that unitarity of the $S$ matrix requires the threshold to be a branch point of the scattering amplitude. Therefore, the square of amplitude modulus always shows a cusp at the threshold and a nontrivial structure may appear in the line shape of the invariant mass distribution. The detailed structure of the cusp, a peak, a dip or buried in the background, depends on the interaction of the two relevant particles near the threshold. More specifically, the cusp at the threshold of a heavy-antiheavy hadron pair will generally show up as a peak in the invariant mass distribution of a heavy quarkonium and light hadron(s), which couple to the heavy-antiheavy hadron pair, if the interaction is attractive at threshold and a nearby virtual state comes into existence (for subtlety and more detailed discussions, we refer to Ref.~\cite{Dong:2020hxe}). If the attraction is strong and a bound state is formed, the peak will be located at the mass of the bound state below threshold. Therefore, the potentials and the pole positions obtained in this work are of great significance for the study of near-threshold line shapes. 

Note that the leading order interaction, which is just a constant, between a heavy-antiheavy hadron pair, when resummed, works well for both purposes mentioned above if we only focus on the near-threshold pole and line shape. Therefore, we only keep the leading order interaction, which is saturated by the vector meson exchange as discussed in the following, and its resummation.

This paper is organized as follows. In Section~\ref{sec:2}, the Lagrangians of heavy hadron and light meson coupling are presented. In Section~\ref{sec:3}, the potentials of different systems are obtained. We calculate the pole positions of different systems and compare them with experimental and other theoretical results in Section~\ref{sec:4}. Section~\ref{sec:5} is devoted to a brief summary. Some details regarding the potentials are relegated to Appendices~\ref{app:vertex}, \ref{app:poten} and \ref{app:cross}.

\section{Lagrangian from heavy quark spin symmetry}
\label{sec:2}

For hadrons containing one or more heavy quarks, additional symmetries emerge in the low energy effective field theory for QCD due to the large heavy quark mass~\cite{Isgur:1989ed,Isgur:1989vq}. On the one hand, the strong suppression of the chromomagnetic interaction, which is proportional to ${\bm\sigma\cdot\bm B}/{m_Q}\sim {\Lambda_{\rm QCD}}/{m_Q}$ with $\Lambda_{\rm QCD}\sim 200-300$ MeV the typical QCD scale, implies that the spin of the heavy quark $s_Q$ is decoupled with the angular momentum $s_\ell$ of light quarks  in the limit of $m_Q\to\infty$. Therefore, HQSS emerges, which means that the interaction is invariant under a transformation of $s_Q$. On the other hand, the change of the velocity of the heavy quark in a singly-heavy hadron during the interaction, $\Delta v={\Delta p}/{m_Q}\sim {\Lambda_{\rm QCD}}/{m_Q}$, vanishes in the limit of $m_Q\to\infty$, and the heavy quark behaves like a static color source, independent of the quark flavor. Therefore, it is expected that the potentials between bottomed hadron pairs are the same as those of the charmed ones, and in turn it is sufficient to focus on the charm sector.

\subsection{Heavy mesons}

To construct a Lagrangian that is invariant under the heavy quark spin transformation and chiral transformation, it is convenient to represent the ground states of charmed mesons as the following superfield~\cite{Wise:1992hn,Casalbuoni:1992gi,Casalbuoni:1996pg}
\begin{align}
H_{a}^{(Q)}=\frac{1+\slashed v}{2}\left[P_{a}^{*(Q) \mu} \gamma_{\mu}-P_{a}^{(Q)} \gamma_{5}\right],
\end{align}
where $a$ is the SU(3) flavor index,
\begin{align}
P^{(Q)}=(D^0,D^+,D_s^+), \quad
P^{*(Q)}_{\mu}=(D^{*0}_{\mu},D^{*+}_\mu,D^{*+}_{s\mu}),
\end{align}
and $v^\mu=p^\mu/M$ is the four-velocity of the heavy meson satisfying $v\cdot v=1$. The heavy field operators contain a factor $\sqrt{M_H}$ and have dimension 3/2. The superfield that creates heavy mesons is constructed as 
\begin{equation}
\bar{H}_{a}^{(Q)}=\gamma_{0} H_{a}^{(Q) \dagger} \gamma_{0}.\label{eq:Hbar}
\end{equation}
The superfields that annihilate or create mesons containing an antiheavy quark are not $\bar{H}_{a}^{(Q)}$ or $H_{a}^{(Q)}$ but the following ones~\cite{Grinstein:1992qt}:
\begin{align}
H_{a}^{(\bar{Q})}&=C\left(\mathcal{C} H_{a}^{(Q)} \mathcal{C}^{-1}\right)^{T} C^{-1}\notag\\
&=\left[P_{a\mu}^{*(\bar{Q})}  \gamma^{\mu}-P_{a}^{(\bar{Q})} \gamma_{5}\right] \frac{1-\slashed{v}}{2},\label{eq:HQbar}\\
\bar{H}_{a}^{(\bar Q)}&=\gamma_{0} H_{a}^{(\bar Q) \dagger} \gamma_{0},\label{eq:HQbarbar}
\end{align}
with 
\begin{align}
P^{(\bar Q)}=(\bar D^0,D^-,D_s^-), \quad
P_\mu^{*(\bar Q)}=(\bar D_\mu^{*0}, D_\mu^{*-}, D^{*-}_{s\mu}).
\end{align}
$\mathcal C$ is the charge conjugation operator and $C=i\gamma^2\gamma^0$ is the charge conjugation matrix, where we have taken the phase convention for charge conjugation as $\mathcal{C} P_{a}^{(Q)} \mathcal{C}^{-1} = P_a^{(\bar Q)}$ and $\mathcal{C} P_{a}^{*(Q)} \mathcal{C}^{-1} = -P_a^{*(\bar Q)}$.

The $P$-wave heavy mesons have two spin multiplets, one with $s_\ell=1/2$ represented by $S$ while the other with $s_\ell=3/2$ represented by $T$~\cite{Falk:1991nq,Falk:1992cx},
\begin{align}
S_{a}^{(Q)}=&\,\frac{1+\slashed v}{2}\left[P_{1 a}^{\prime(Q) \mu} \gamma_{\mu} \gamma_{5}-P_{0 a}^{*(Q)}\right],\\
T_{a}^{(Q) \mu}=&\,\frac{1+\slashed v}{2}\bigg[P_{2 a}^{*(Q) \mu \nu} \gamma_{\nu}\notag\\
&-\sqrt{\frac{3}{2}} P_{1 a \nu}^{(Q)} \gamma_{5}\left(g^{\mu \nu}-\frac{1}{3} \gamma^{\nu}\left(\gamma^{\mu}-v^{\mu}\right)\right)\bigg].
\end{align}
Analogous with Eqs.~(\ref{eq:Hbar},\ref{eq:HQbar},\ref{eq:HQbarbar}), we have
\begin{align}
\bar{S}_{a}^{(Q)}=&\,\gamma_{0} S_{a}^{(Q) \dagger} \gamma_{0},\\
\bar{T}_{a}^{(Q) \mu}=&\, \gamma_{0} T_{a}^{(Q) \mu \dagger} \gamma_{0},\\
S_{a}^{(\bar{Q})}=&\,\left[P_{1 a}^{\prime(\bar{Q}) \mu} \gamma_{\mu} \gamma_{5}-P_{0 a}^{*(\bar{Q})}\right] \frac{1-\slashed v}{2},\\
T_{a}^{(\bar{Q}) \mu}=&\,\left[P_{2 a}^{(\bar{Q}) \mu \nu} \gamma_{\nu}-\sqrt{\frac{3}{2}} P_{1 a \nu}^{(\bar{Q})}\right.\notag\\
&\left.\times \gamma_{5}\left(g^{\mu \nu}-\frac{1}{3}\left(\gamma^{\mu}-v^{\mu}\right) \gamma^{\nu}\right)\right] \frac{1-\slashed v}{2},\\
\bar{S}_{a}^{(\bar{Q})}=&\,\gamma_{0} S_{a}^{(\bar{Q}) \dagger} \gamma_{0},\\
 \bar{T}_{a \mu}^{(\bar{Q})}=&\,\gamma_{0} T_{a \mu}^{(\bar{Q}) \dagger} \gamma_{0}.
\end{align}
The mesons in the $T$ multiplet are
\begin{align}
P_1^{(Q)}&=(D_1(2420)^0,D_1(2420)^+,D_{s1}(2536)^+), \notag\\
P_2^{(Q)}&=(D_2(2460)^0,D_2(2460)^+,D_{s2}(2573)^+),
\end{align}
which can couple to $D^*\pi/K$ only in $D$-wave in the heavy quark limit. 
While the $P$-wave charmed mesons with $s_\ell=1/2$ can couple to $D^*\pi/K$ in $S$-wave without violating HQSS, there are issues in identifying them.
The $D_0^*(2300)$ and $D_1(2430)$ listed in the Review of Particle Physics (RPP)~\cite{Zyla:2020zbs} could be candidates for the charm-nonstrange ones. However, on the one hand, they have rather large widths such that they would have decayed before they can be bound together with another heavy hadron~\cite{Filin:2010se,Guo:2011dd}; on the other hand, they were extracted using the Breit-Wigner parameterization which has deficiencies in the current case~\cite{Du:2019oki} and has been demonstrated~\cite{Du:2020pui} to lead to resonance parameters for the $D_0^*(2300)$ in conflict with the precise LHCb data of the $B^-\to D^+\pi^-\pi^-$ process~\cite{Aaij:2016fma}.
For the ones with strangeness, the lowest positive-parity $D_{s0}^*(2317)$ and $D_{s1}(2460)$ are widely considered as molecular states of $DK$ and $D^*K$~\cite{Barnes:2003dj,vanBeveren:2003kd,Kolomeitsev:2003ac,Chen:2004dy,Guo:2006fu,Guo:2006rp}, see Ref.~\cite{Guo:2019dpg} for a recent review collecting evidence for such an interpretation. 
This multiplet $S$, therefore, will not be considered in the rest of this work. For studies of three-body hadronic molecular states involving the $D_{s0}^*(2317)$ as a $DK$ subsystem, we refer to Refs.~\cite{Ma:2017ery,MartinezTorres:2018zbl,Wu:2019vsy,Wu:2020job}.

The light pseudoscalar meson octet can be introduced using the nonlinear realization of the spontaneous chiral symmetry breaking of QCD as $\Sigma=\xi^2$ and $\xi=e^{i\Pi/(\sqrt{2}F_\pi)}$ with $F_\pi=92$~MeV the pion decay constant and 
\begin{align}
\Pi=\left(\begin{array}{ccc}
\frac{\pi^{0}}{\sqrt{2}}+\frac{\eta}{\sqrt{6}} & \pi^{+} & K^{+} \\
\pi^{-} & -\frac{\pi^{0}}{\sqrt{2}}+\frac{\eta}{\sqrt{6}} & K^{0} \\
K^{-} & \bar{K}^{0} & -\sqrt{\frac{2}{3}} \eta
\end{array}\right).
\end{align}
The effective Lagrangian for the coupling of heavy mesons and light pseudoscalar mesons is constructed by imposing invariance under both heavy quark spin transformation and chiral transformation~\cite{Wise:1992hn,Falk:1992cx,Grinstein:1992qt},
\begin{widetext}
\begin{align}
\mathcal{L}_{PP\Pi}&=\,i g\left\langle H_{b}^{(Q)} \slashed{\mathcal{A}}_{b a} \gamma_{5} \bar{H}_{a}^{(Q)}\right\rangle+i k\left\langle T_{b}^{(Q) \mu} \slashed{\mathcal{A}}_{b a} \gamma_{5} \bar{T}_{a \mu}^{(Q)}\right\rangle+i \tilde{k}\left\langle S_{b}^{(Q)} \slashed{\mathcal{A}}_{b a} \gamma_{5} \bar{S}_{a}^{(Q)}\right\rangle+\left[i h\left\langle S_{b}^{(Q)} \slashed{\mathcal{A}}_{b a} \gamma_{5} \bar{H}_{a}^{(Q)}\right\rangle\right.\notag \\
&\left.+\,i \tilde{h}\left\langle T_{b}^{(Q) \mu} \mathcal{A}_{\mu b a} \gamma_{5} \bar{S}_{a}^{(Q)}\right\rangle+i \frac{h_{1}}{\Lambda_{\chi}}\left\langle T_{b}^{(Q) \mu}\left(D_{\mu} \slashed{\mathcal{A}}\right)_{b a} \gamma_{5} \bar{H}_{a}^{(Q)}\right\rangle+i \frac{h_{2}}{\Lambda_{\chi}}\left\langle T_{b}^{(Q) \mu}\left(\slashed D \mathcal{A}_{\mu}\right)_{b a} \gamma_{5} \bar{H}_{a}^{(Q)}\right\rangle+h . c .\right]\notag\\
&+i g\left\langle\bar{H}_{a}^{(\bar{Q})} \slashed{\mathcal{A}}_{a b} \gamma_{5} H_{b}^{(\bar{Q})}\right\rangle+i k\left\langle\bar{T}_{a}^{(\bar{Q}) \mu} \slashed{\mathcal{A}}_{a b} \gamma_{5} T_{b \mu}^{(\bar{Q})}\right\rangle+i \tilde{k}\left\langle\bar{S}_{a}^{(\bar{Q})} \slashed{\mathcal{A}}_{a b} \gamma_{5} S_{b}^{(\bar{Q})}\right\rangle+{\bigg[}i h\left\langle\bar{H}_{a}^{(\bar{Q})} \slashed{\mathcal{A}}_{a b} \gamma_{5} S_{b}^{(\bar{Q})}\right\rangle\notag\\
&+i \tilde{h}\left\langle\bar{S}_{a}^{(\bar{Q})} \mathcal{A}_{\mu a b} \gamma_{5} T_{b}^{(\bar{Q}) \mu}\right\rangle+i \frac{h_{1}}{\Lambda_{\chi}}\left\langle\bar{H}_{a}^{(\bar{Q})}\Big(\slashed{\mathcal{A}} \stackrel{\leftarrow}{D_{\mu}^{\prime}}\Big)_{a b} \gamma_{5} T_{b}^{(\bar{Q}) \mu}\right\rangle+i \frac{h_{2}}{\Lambda_{\chi}}\left\langle\bar{H}_{a}^{(\bar{Q})}\Big(\mathcal{A}_{\mu} \stackrel{\leftarrow}{\slashed{D}^{\prime}}\Big)_{a b} \gamma_{5} T_{b}^{(\bar{Q}) \mu}\right\rangle+\text {h.c.}{\bigg]},
\label{eq:DDP}
\end{align}
\end{widetext}
where $D_{\mu}=\partial_{\mu}+\mathcal{V}_{\mu}$, $D'_{\mu}=\partial_{\mu}-\mathcal{V}_{\mu}$, $\langle\cdots\rangle$ denotes tracing over the Dirac $\gamma$ matrices, $\Lambda_\chi\simeq 4\pi F_\pi$ is the chiral symmetry breaking scale, and
\begin{align}
\mathcal{V}_{\mu}&=\frac{1}{2}\left(\xi^{\dagger} \partial_{\mu} \xi+\xi \partial_{\mu} \xi^{\dagger}\right),\\
\mathcal{A}_{\mu}&=\frac{1}{2}\left(\xi^{\dagger} \partial_{\mu} \xi-\xi \partial_{\mu} \xi^{\dagger}\right)
\end{align}
are the vector and axial currents which contain an even and odd number of pseudoscalar mesons, respectively. 

The coupling of heavy mesons and light vector mesons can be introduced by using the hidden local symmetry approach~\cite{Bando:1984ej,Bando:1987br,Meissner:1987ge}, and the Lagrangian reads~\cite{Casalbuoni:1992gi,Casalbuoni:1992dx,Casalbuoni:1996pg}
\begin{widetext}
\begin{align}
\mathcal{L}_{PPV}=&\,i \beta\left\langle H_{b}^{(Q)} v^{\mu}\left(\mathcal{V}_{\mu}-\rho_{\mu}\right)_{b a} \bar{H}_{a}^{(Q)}\right\rangle+i \lambda\left\langle H_{b}^{(Q)} \sigma^{\mu \nu} F_{\mu \nu}(\rho)_{b a} \bar{H}_{a}^{(Q)}\right\rangle+i \beta_{1}\left\langle S_{b}^{(Q)} v^{\mu}\left(\mathcal{V}_{\mu}-\rho_{\mu}\right)_{b a} \bar{S}_{a}^{(Q)}\right\rangle\notag\\
&+i \lambda_{1}\left\langle S_{b}^{(Q)} \sigma^{\mu \nu} F_{\mu \nu}(\rho)_{b a} \bar{S}_{a}^{(Q)}\right\rangle+i \beta_{2}\left\langle T_{b}^{(Q) \lambda} v^{\mu}\left(\mathcal{V}_{\mu}-\rho_{\mu}\right)_{b a} \bar{T}_{a \lambda}^{(Q)}\right\rangle+i \lambda_{2}\left\langle T_{b}^{(Q) \lambda} \sigma^{\mu \nu} F_{\mu \nu}(\rho)_{b a} \bar{T}_{a \lambda}^{(Q)}\right\rangle\notag\\
&+\left[i \zeta\left\langle H_{b}^{(Q)} \gamma^{\mu}\left(\mathcal{V}_{\mu}-\rho_{\mu}\right)_{b a} \bar{S}_{a}^{(Q)}\right\rangle+i \mu\left\langle H_{b}^{(Q)} \sigma^{\lambda \nu} F_{\lambda \nu}(\rho)_{b a} \bar{S}_{a}^{(Q)}\right\rangle+i \zeta_{1}\left\langle T_{b}^{(Q) \mu}\left(\mathcal{V}_{\mu}-\rho_{\mu}\right)_{b a} \bar{H}_{a}^{(Q)}\right\rangle\right.\notag\\
&\left.+\mu_{1}\left\langle T_{b}^{(Q) \mu} \gamma^{\nu} F_{\mu \nu}(\rho)_{b a} \bar{H}_{a}^{(Q)}\right\rangle+h . c .\right]\notag\\
&-i \beta\left\langle\bar{H}_{a}^{(\bar{Q})} v^{\mu}\left(\mathcal{V}_{\mu}-\rho_{\mu}\right)_{a b} H_{b}^{(\bar{Q})}\right\rangle+i \lambda\left\langle\bar{H}_{a}^{(\bar{Q})} \sigma^{\mu \nu} F_{\mu \nu}(\rho)_{a b} H_{b}^{(\bar{Q})}\right\rangle-i \beta_{1}\left\langle\bar{S}_{a}^{(\bar{Q})} v^{\mu}\left(\mathcal{V}_{\mu}-\rho_{\mu}\right)_{a b} S_{b}^{(\bar{Q})}\right\rangle\notag\\
&+i \lambda_{1}\left\langle\bar{S}_{a}^{(\bar{Q})} \sigma^{\mu \nu} F_{\mu \nu}(\rho)_{a b} S_{b}^{(\bar{Q})}\right\rangle-i \beta_{2}\left\langle\bar{T}_{a \lambda}^{(\bar{Q})} v^{\mu}\left(\mathcal{V}_{\mu}-\rho_{\mu}\right)_{a b} T_{b}^{(\bar{Q}) \lambda}\right\rangle+i \lambda_{2}\left\langle\bar{T}_{a \lambda}^{(\bar{Q})} \sigma^{\mu \nu} F_{\mu \nu}(\rho)_{a b} T_{b}^{(\bar{Q}) \lambda}\right\rangle\notag\\
&+\left[i \zeta\left\langle\bar{S}_{a}^{(\bar{Q})} \gamma^{\mu}\left(\mathcal{V}_{\mu}-\rho_{\mu}\right)_{a b} H_{a}^{(\bar{Q})}\right\rangle+i \mu\left\langle\bar{S}_{a}^{(\bar{Q})} \sigma^{\lambda \nu} F_{\lambda \nu}(\rho)_{a b} H_{b}^{(\bar{Q})}\right\rangle-i \zeta_{1}\left\langle\bar{H}_{a}^{(\bar{Q})}\left(\mathcal{V}_{\mu}-\rho_{\mu}\right)_{a b} T_{b}^{(\bar{Q}) \mu}\right\rangle\right.\notag\\
&\left.+\mu_{1}\left\langle\bar{H}_{a}^{(\bar{Q})} \gamma^{\nu} F_{\mu \nu}(\rho)_{a b} T_{b}^{(\bar{Q}) \mu}\right\rangle+\text {h.c.}\right],\label{eq:DDV}
\end{align}
\end{widetext}
with $F_{\mu \nu}=\partial_{\mu} \rho_{\nu}-\partial_{\nu} \rho_{\mu}+\left[\rho_{\mu}, \rho_{\nu}\right]$, and
\begin{align}
\rho=i \frac{g_{V}}{\sqrt{2}} V=i \frac{g_{V}}{\sqrt{2}}\left(\begin{array}{ccc}
\frac{\omega}{\sqrt{2}}+\frac{\rho^{0}}{\sqrt{2}} & \rho^{+} & K^{*+} \\
\rho^{-} & \frac{\omega}{\sqrt{2}}-\frac{\rho^{0}}{\sqrt{2}} & K^{* 0} \\
K^{*-} & \bar{K}^{* 0} & \phi
\end{array}\right),
\end{align}
which satisfies $\mathcal C V \mathcal C^{-1}=-V^T$. 

Remind that in the following we are only interested in the potential near threshold and will not consider coupled channels. Therefore, the Lagrangian that results in potentials proportional to the transferred momentum $\bm q^2$ will have little contributions. At the leading order of the chiral expansion, the light pseudoscalar mesons as Goldstone bosons only couple in derivatives, as demonstrated in Eq.~(\ref{eq:DDP}), so all pseudoscalar exchanges have subleading contributions near threshold in comparison with the constant contact term that can generate a near-threshold pole after resummation. 
Moreover, coupled channels are not taken into account here, and we do not consider the $s_\ell=1/2$ mesons. To this end we can just keep the $\beta$, $\beta_2$ and $\zeta_1$ terms in Eq.~(\ref{eq:DDV}). Expanding these terms we obtain
\begin{align}
&\mathcal L_{PPV}=-\sqrt{2}\beta g_V \left(P_a^{(Q)}P_b^{(Q)\dagger}-P_b^{(\bar Q)}P_a^{(\bar Q)\dagger}\right)v_\mu V^\mu_{ab}\notag\\
&+\sqrt{2}\beta g_V \left(P_a^{*(Q)\nu}P_{b\nu}^{*(Q)\dagger}-P_b^{*(\bar Q)\nu}P_{a\nu}^{*(\bar Q)\dagger}\right)v_\mu V^\mu_{ab}\notag\\
&-\sqrt{2}\beta_2 g_V \left(P_{1a}^{(Q)\nu}P_{1b\nu}^{(Q)\dagger}-P_{1b}^{(\bar Q)\nu}P_{1a\nu}^{(\bar Q)\dagger}\right)v_\mu V^\mu_{ab}\notag\\
&+\sqrt{2}\beta_2 g_V \left(P_{2a}^{(Q)\alpha\beta}P_{2b\alpha\beta}^{(Q)\dagger}-P_{2b}^{(\bar Q)\alpha\beta}P_{2a\alpha\beta}^{(\bar Q)\dagger}\right)v_\mu V^\mu_{ab}\notag\\
&+\Big[\sqrt2\zeta_1 g_V \left(P_{2a}^{(Q)\mu\nu}P_{b\nu}^{(Q)*\dagger}+P_{2b}^{(\bar Q)\mu\nu}P_{a\nu}^{(\bar Q)*\dagger}\right)V_{ab\mu}\notag\\
&-\frac{i\zeta_1 g_V}{\sqrt3} \epsilon_{\alpha\beta\gamma\delta}\left(P_{1a}^{(Q)\alpha} P_b^{(Q)*\dagger\beta}+P_{1b}^{(\bar Q)\alpha}  P_a^{(\bar Q)*\dagger\beta}\right)v^\gamma V_{ab}^\delta\notag\\
&-\frac{2\zeta_1 g_V}{\sqrt3} \left(P_{1a\mu}^{(Q)}P_b^{(Q)\dagger}-P_{1b\mu}^{(\bar Q) }P_a^{(\bar Q)\dagger}\right) V_{ab}^\mu+{\rm h.c.}\Big].\label{eq:LPPV}
\end{align}
Assuming vector meson dominance, the coupling constants $g_V$ and $\beta$ were estimated to be $5.8$~\cite{Bando:1987br} and $0.9$~\cite{Isola:2003fh}, respectively. In Ref.~\cite{Dong:2019ofp}, it is estimated that $\beta_2\approx-\beta=-0.9$ under the assumption that the coupling of $D_1D_1V$ is the same as that of $DDV$ and $\zeta_1\approx0.16$ from the decay of $K_1\to K\rho$. 

\subsection{Heavy baryons}

In the heavy quark limit, the ground states of heavy baryons $Qqq$ form an SU(3) antitriplet with $J^P = \frac12^+$ denoted by $B^{(Q)}_{\bar 3}$ and two degenerate sextets with $J^P = (\frac12,\frac32)^+$ denoted by $(B^{(Q)}_{6},B^{(Q)*}_6)$~\cite{Yan:1992gz},
\begin{align}
B^{(Q)}_{\bar{3}}&=\left(\begin{array}{ccc}
0 & \Lambda_{c}^{+} & \Xi_{c}^{+} \\
-\Lambda_{c}^{+} & 0 & \Xi_{c}^{0} \\
-\Xi_{c}^{+} & -\Xi_{c}^{0} & 0
\end{array}\right),
\\ B^{(Q)}_{6}&=\left(\begin{array}{ccc}
\Sigma_{c}^{++} & \frac{1}{\sqrt{2}} \Sigma_{c}^{+} & \frac{1}{\sqrt{2}} \Xi_{c}^{\prime+} \\
\frac{1}{\sqrt{2}} \Sigma_{c}^{+} & \Sigma_{c}^{0} & \frac{1}{\sqrt{2}} \Xi_{c}^{\prime 0} \\
\frac{1}{\sqrt{2}} \Xi_{c}^{\prime+} & \frac{1}{\sqrt{2}} \Xi_{c}^{\prime 0} & \Omega_{c}^{0}
\end{array}\right), \\
B_{6}^{(Q)*}&=\left(\begin{array}{ccc}
\Sigma_{c}^{*++} & \frac{1}{\sqrt{2}} \Sigma_{c}^{*+} & \frac{1}{\sqrt{2}} \Xi_{c}^{*+} \\
\frac{1}{\sqrt{2}} \Sigma_{c}^{*+} & \Sigma_{c}^{* 0} & \frac{1}{\sqrt{2}} \Xi_{c}^{* 0} \\
\frac{1}{\sqrt{2}} \Xi_{c}^{*+} & \frac{1}{\sqrt{2}} \Xi_{c}^{* 0} & \Omega_{c}^{* 0}
\end{array}\right).
\end{align}
Here we do not consider the $P$-wave heavy baryons since they are not well established experimentally. The two sextets are collected into the superfield $S_\mu$,
\begin{align}
S^{(Q)}_{\mu}&=B_{6 \mu}^{(Q)*}- \frac{1}{\sqrt{3}}\left(\gamma_{\mu}+v_{\mu}\right) \gamma^{5} B^{(Q)}_{6},\\
\bar S^{(Q)}_{\mu}&=\bar B_{6 \mu}^{(Q)*}+ \frac{1}{\sqrt{3}}\bar B^{(Q)}_6\gamma^{5} \left(\gamma_{\mu}+v_{\mu}\right),
\end{align}
where $B_{6\mu}$ is the Rarita-Schwinger vector-spinor field~\cite{Rarita:1941mf}. The fields that annihilate anti-baryons are obtained by taking the charge conjugation of  $B^{(Q)}_{\bar 3}$, $B^{(Q)}_6$ and $B_6^{(Q)*}$,
\begin{align}
B^{(\bar Q)}_{{3}}&=\left(\begin{array}{ccc}
0 & \Lambda_{c}^{-} & \Xi_{c}^{-} \\
-\Lambda_{c}^{-} & 0 & \bar\Xi_{c}^{0} \\
-\Xi_{c}^{-} &- \bar\Xi_{c}^{0} & 0
\end{array}\right), \\
B^{(\bar Q)}_{\bar6}&=\left(\begin{array}{ccc}
\Sigma_{c}^{--} & \frac{1}{\sqrt{2}} \Sigma_{c}^{-} & \frac{1}{\sqrt{2}} \Xi_{c}^{\prime-} \\
\frac{1}{\sqrt{2}} \Sigma_{c}^{-} & \bar\Sigma_{c}^{0} & \frac{1}{\sqrt{2}} \bar\Xi_{c}^{\prime 0} \\
\frac{1}{\sqrt{2}} \Xi_{c}^{\prime-} & \frac{1}{\sqrt{2}} \bar\Xi_{c}^{\prime 0} & \bar\Omega_{c}^{0}
\end{array}\right), \\
B_{\bar6}^{(\bar Q)*}&=\left(\begin{array}{ccc}
\Sigma_{c}^{*--} & \frac{1}{\sqrt{2}} \Sigma_{c}^{*-} & \frac{1}{\sqrt{2}} \Xi_{c}^{*-} \\
\frac{1}{\sqrt{2}} \Sigma_{c}^{*-} & \bar\Sigma_{c}^{* 0} & \frac{1}{\sqrt{2}} \bar\Xi_{c}^{* 0} \\
\frac{1}{\sqrt{2}} \Xi_{c}^{*-} & \frac{1}{\sqrt{2}} \bar\Xi_{c}^{* 0} & \bar\Omega_{c}^{* 0}
\end{array}\right),
\end{align}
where we have used the phase conventions such that $\mathcal C B^{(Q)} \mathcal C^{-1}=B^{(\bar Q)}$. The corresponding superfields now read
\begin{align}
S^{(\bar Q)}_{\mu}&=B_{\bar6 \mu}^{(\bar Q)*}- \frac{1}{\sqrt{3}}\left(\gamma_{\mu}+v_{\mu}\right) \gamma^{5} B^{(\bar Q)}_{\bar6},\\
\bar S^{(\bar Q)}_{\mu}&=\bar B_{\bar6 \mu}^{(\bar Q)*}+ \frac{1}{\sqrt{3}}\bar B^{(\bar Q)}_{\bar6 }\gamma^{5} \left(\gamma_{\mu}+v_{\mu}\right).
\end{align}

The Lagrangian for the coupling of heavy baryons and light mesons is constructed as~\cite{Liu:2011xc}
\begin{align}
\mathcal{L}_{\mathcal{B}}=&\, \mathcal{L}_{B_{3}}+\mathcal{L}_{S}+\mathcal{L}_{\text {int}}, \\
\mathcal{L}_{B_{{\bar 3}}}=&\, \frac{1}{2} \operatorname{tr}\left[\bar{B}^{(Q)}_{\bar{3}}(i v \cdot D) B^{(Q)}_{\bar{3}}\right]\notag\\
&+i \beta_{B} \operatorname{tr} \left[\bar{B}^{(Q)}_{\bar{3}} v^{\mu}\left(\mathcal V_{\mu}-\rho_{\mu}\right) B^{(Q)}_{\bar{3}}\right],\\
\mathcal{L}_{S}=&\,-\operatorname{tr}\left[\bar{S}^{(Q)\alpha}\left(i v \cdot D-\Delta_{B}\right) S^{(Q)}_{\alpha}\right]\notag\\
&+\frac{3}{2} g_{1}\left(i v_{\kappa}\right) \epsilon^{\mu \nu \lambda \kappa} \operatorname{tr}\left[\bar{S}^{(Q)}_{\mu} \mathcal A_{\nu} S^{(Q)}_{\lambda}\right] \notag\\
&+i \beta_{S} \operatorname{tr}\left[\bar{S}^{(Q)}_{\mu} v_{\alpha}\left(\mathcal V^{\alpha}-\rho^{\alpha}\right) S^{(Q)\mu}\right]\notag\\
&+\lambda_{S} \operatorname{tr}\left[\bar{S}^{(Q)}_{\mu} F^{\mu \nu} S^{(Q)}_{\nu}\right],\\
\mathcal{L}_{\rm i n t}=&\, g_{4} \operatorname{tr}\left[\bar{S}^{(Q)\mu}\mathcal A_{\mu} B^{(Q)}_{\bar{3}}\right]\notag\\
&+i \lambda_{I} \epsilon^{\mu \nu \lambda \kappa} v_{\mu} \operatorname{tr}\left[\bar{S}^{(Q)}_{\nu} F_{\lambda \kappa} B^{(Q)}_{\bar{3}}\right]+{\rm h.c.},
\end{align}
where $D_\mu B=\partial_\mu B+\mathcal V_\mu B+B\mathcal V_\mu^T$, and $\Delta_B=m_{6}-m_{\bar3}$ is the mass difference between the anti-triplet and sextet baryons.  The coupling constants $\beta_B$ and $\beta_S$ are estimated in Ref.~\cite{Liu:2011xc} where $\beta_S=-2\beta_B=1.44$ or $2.06$ from quark model and  $\beta_S=-2\beta_B=1.74$ from the vector meson dominance assumption. However, there is a sign ambiguity (only the absolute value of the $\rho NN$ coupling was determined from the $NN$ scattering~\cite{Machleidt:1987hj}). It turns out that the sign choice in Ref.~\cite{Liu:2011xc} yields potentials of the anti-charmed meson and charmed baryon systems with an opposite sign compared to the ones obtained by SU(4) relations~\cite{Wu:2010vk}. It also conflicts with the famous $P_c$ states~\cite{Aaij:2019vzc}, which are believed to be molecular states of $\bar D^{(*)}\Sigma_c^{(*)}$ with isospin $1/2$---with a positive $\beta_S$, these systems will be repulsive (see below). These issues can be fixed by choosing the signs of $\beta_B$ and $\beta_S$ opposite to those taken in Ref.~\cite{Liu:2011xc}, just like what Ref.~\cite{Chen:2019asm} did. 

For the coupling of antiheavy baryons and light mesons, by taking the charge conjugation transformation of the above ones, we have
\begin{align}
\mathcal{L}'_{\mathcal{B}}=&\, \mathcal{L}'_{B_{3}}+\mathcal{L}'_{S}+\mathcal{L}'_{\text {int}}, \\
\mathcal{L}'_{B_{{\bar 3}}}=&\, \frac{1}{2} \operatorname{tr}\left[\bar{B}^{(\bar Q)}_{{3}}(i v \cdot D)^T B^{(\bar Q)}_{{3}}\right]\notag\\
&-i \beta_{B} \operatorname{tr}\left[\bar{B}^{(\bar Q)}_{{3}} v^{\mu}\left(\mathcal V_{\mu}-\rho_{\mu}\right)^T B^{(\bar Q)}_{{3}}\right],\\
\mathcal{L}'_{S}=&\,-\operatorname{tr}\left[\bar{S}^{(\bar Q)\alpha}\left(i v \cdot D-\Delta_{B}\right)^T S^{(\bar Q)}_{\alpha}\right]\notag\\
&+\frac{3}{2} g_{1}\left(i v_{\kappa}\right) \epsilon^{\mu \nu \lambda \kappa} \operatorname{tr}\left[\bar{S}^{(\bar Q)}_{\mu} \mathcal A_{\nu}^T S^{(\bar Q)}_{\lambda}\right]\notag \\
&-i \beta_{S} \operatorname{tr}\left[\bar{S}^{(\bar Q)}_{\mu} v_{\alpha}\left(\mathcal V^{\alpha}-\rho^{\alpha}\right)^T S^{(\bar Q)\mu}\right]\notag\\
&+\lambda_{S} \operatorname{tr}\left[\bar{S}^{(\bar Q)}_{\mu} (F^{\mu \nu})^T S^{(\bar Q)}_{\nu}\right],\\
\mathcal{L}'_{\rm int}=&\, g_{4} \operatorname{tr}\left[\bar{S}^{(\bar Q)\mu}\mathcal A^T_{\mu} B^{(\bar Q)}_{{3}}\right]\notag\\
&+i \lambda_{I} \epsilon^{\mu \nu \lambda \kappa} v_{\mu} \operatorname{tr}\left[\bar{S}^T_{\nu}( F_{\lambda \kappa})^T B^{(\bar Q)}_{{3}}\right]+ {\rm h.c.},
\end{align}
with the transpose acting on the SU(3) flavor matrix. Notice that the spinor for an antibaryon is $u$ instead of $v$ since here the fields of heavy baryons and heavy antibaryons are treated independently.

Similar with the Lagrangian for heavy mesons, we can focus on the vector exchange contributions, and only the following terms are relevant,
\begin{align}
\mathcal{L}_{BBV}&=\, i \beta_{B} \mathrm{tr}\left[\bar{B}^{(Q)}_{\bar{3}} v^{\mu}\left(\mathcal V_{\mu}-\rho_{\mu}\right) B^{(Q)}_{\bar{3}}\right]\notag\\
&-i \beta_{B}\mathrm{tr}\left[\bar{B}^{(\bar Q)}_{{3}} v^{\mu}\left(\mathcal V_{\mu}-\rho_{\mu}\right)^T B^{(\bar Q)}_{{3}}\right]\notag\\
&+ i \beta_{S} \operatorname{tr}\left[\bar{S}_{\nu}^{(Q)} v^{\mu}\left(\mathcal V_{\mu}-\rho_{\mu}\right) S^{(Q)\nu}\right]\notag\\
&-i \beta_{S}\operatorname{tr}\left[\bar{S}_{\nu}^{(\bar Q)} v^{\mu}\left(\mathcal V_{\mu}-\rho_{\mu}\right)^T S^{(\bar Q)\nu}\right].\label{eq:LSSV}
\end{align}

\section{Potentials}
\label{sec:3}

\subsection{Conventions}

In this paper, we take the following charge conjugation conventions: 
\begin{align}
\mathcal C \ket{D}& =\ket{\bar D}, & \mathcal C\ket{D^*}&=-\ket{\bar D^*}, \notag\\ \mathcal C \ket{D_1}&=\ket{\bar D_1},& \mathcal C\ket{D_2}&=-\ket{\bar D_2}, \notag\\ 
\mathcal C \ket{B_{\bar 3}}&=\ket{\bar B_{\bar 3}}, &\mathcal C \ket{B^{(*)}_{6}}&=\ket{\bar B^{(*)}_{6}},
\end{align} 
which are consistent with the Lagrangians in Section~\ref{sec:2}. Within these conventions, the flavor wave functions of the flavor-neutral systems that are charge conjugation eigenstates, including $\ket{D D^*}_c$, $\ket{D D_1}_c$, $\ket{D^*D_1}_c$, $\ket{D D_2}_c$, $\ket{D^*D_2}_c$, $\ket{D_1D_2}_c$ and $\ket{B_6B_{ 6}^{*}}_c$, can be expressed as
\begin{align}
    \ket{A_1A_2}_c=\frac{1}{\sqrt2}\lra{\ket{A_1\bar A_2} \pm (-1)^{J-J_1-J_2}c\,c_1c_2\ket{A_2\bar A_1}},\label{eq:cpm}
\end{align}
where $J_i$ is the spin of $\ket{A_i}$, $J$ is the total spin of the system $\ket{A_1A_2}_c$, $c_i$ is defined by $\mathcal C \ket{A_i}=c_i\ket{\bar A_i}$, and the plus and minus between the two terms are for boson-boson and fermion-fermion systems, respectively. These systems satisfy $\mathcal C \ket{A_1A_2}_c=c\ket{A_1A_2}_c$ with $c=\pm1$.

For the isospin conventions, we use the following ones:
\begin{align}
\ket{u}&=\ket{\frac12,\frac12},\quad~~\, \ket{d}=\ket{\frac12,-\frac12},\notag\\
 \ket{\bar d}&=\ket{\frac12,\frac12},\quad \ket{\bar u}=-\ket{\frac12,-\frac12}.
\end{align}
Consequently, we have
\begin{align}
\ket{D^{+}}&=\ket{\frac12,\frac12},& \ket{D^{-}}&=\ket{\frac12,-\frac12},\notag\\
\ket{D^{0}}&=-\ket{\frac12,-\frac12},&  \ket{\bar D^{0}}&=\ket{\frac12,\frac12},\notag\\
\ket{D_s^{+}}&=\ket{0,0},& \ket{D_s^{-}}&=\ket{0,0},\notag\\
\ket{\Lambda_c^+}&=\ket{0,0},& \ket{\Lambda_c^-}&=-\ket{0,0},\notag\\
\ket{\Xi_c^{('*)+}}&=\ket{\frac12,\frac12},& \ket{\Xi_c^{('*)-}}&=\ket{\frac12,-\frac12},\notag\\
\ket{\Xi_c^{('*)0}}&=\ket{\frac12,-\frac12},& \ket{\bar\Xi_c^{('*)0}}&=-\ket{\frac12,\frac12},\notag\\
\ket{\Sigma_c^{(*)++}}&=\ket{1,1},& \ket{\Sigma_c^{(*)--}}&=\ket{1,-1},\notag\\
\ket{\Sigma_c^{(*)+}}&=\ket{1,0},& \ket{\Sigma_c^{(*)-}}&=-\ket{1,0},\notag\\
\ket{\Sigma_c^{(*)0}}&=\ket{1,-1},& \ket{\bar\Sigma_c^{(*)0}}&=\ket{1,1},\notag\\
\ket{\Omega_c^{(*)0}}&=\ket{0,0},& \ket{\bar\Omega_c^{(*)0}}&=\ket{0,0}.
\end{align}
The isospin states of $D^*$, $D_1$ and $D_2^*$ are the same as those of $D$. The flavor wave functions of the systems considered below with certain isospins can be easily computed using Clebsch-Gordan coefficients with these conventions.

The potential we calculate is $V=-\mathcal M$ with $\mathcal M$ the $2\to 2$ invariant scattering amplitude so that a negative $V$ means an attraction interaction. This convention is the same as the widely used one in the on-shell Bethe-Salpeter equation $T=V+VGT$~\cite{Oller:1997ti}, and it is also the same as the nonrelativistic potential in the Schr\"odinger equation up to a mass factor. 

\subsection{Potentials from light vector exchange}

With the Lagrangian and conventions presented above we are ready to calculate the potentials of different systems.
We will use the resonance saturation model to get the approximate potentials as constant contact terms. 
Note that the resonance saturation has been known to be able to well approximate the low-energy constants (LECs) in the higher order Lagrangians of chiral perturbation theory~\cite{Ecker:1988te,Donoghue:1988ed}, and it turns out therein that whenever vector mesons contribute they dominate the numerical values of the LECs at the scale around the $\rho$-meson mass, which is called the modern version of vector meson dominance.

The general form of $\mathcal M$ for a process $A_1(p_1)\bar A_2(p_2)\to A_1(k_1)\bar A_2(k_2)$ by the vector meson exchange reads
\begin{align}
    i\mathcal M= ig_1 v_\mu \frac{-i(g^{\mu\nu}-{q^\mu q^\nu}/{m_{\rm ex}^2})}{q^2-m_{\rm ex}^2+i\epsilon} v_\nu ig_2\approx -i\frac{g_1 g_2}{m_{\rm ex}^2},\label{eq:Amp}
\end{align}
where $g_1$ and $g_2$ account for the vertex information for $A_1$ and $\bar A_2$, respectively, $q = p_1-k_1$, and we have neglected terms suppressed by $\mathcal{O}(\vec q\,^2/m_\text{ex}^2)$. For different particles $g_1$ and $g_2$ are collected in Appendix~\ref{app:vertex}.
 It is worth mentioning that the spin information of the component particles is irrelevant here since the exchanged vectors only carry the momentum information, see Eqs.~(\ref{eq:LPPV},\ref{eq:LSSV}). Hence for a given system with different total spins, the potentials at threshold are the same. With the vertex factors evaluated, the potentials of different systems have a uniform expression,
\begin{equation}
    V\approx -F \tilde \beta_1 \tilde \beta_2g_V^2\frac{2m_1m_2}{m_{\rm ex}^2},
    \label{eq:potential}
\end{equation}
where $m_1,m_2$ and $m_{\rm ex}$ are the masses of the two heavy hadrons and the exchanged particle, respectively. $\tilde \beta_1$ and $\tilde \beta_2$ are the coupling constants for the two heavy hadrons with the vector mesons, and, given explicitly, $\tilde \beta_i=\beta$ for the $S$-wave charmed mesons, $\tilde \beta_i=-\beta$ for the $P$-wave charmed mesons,  $\tilde \beta_i=\beta_B$ for the anti-triplet baryons, and  $\tilde \beta_i=-\beta_S/2$ for the sextet baryons.  $F$ is a group theory factor accounting for the light-flavor SU(3) information, and in our convention a positive $F$ means an attractive interaction. The values of $F$ are listed in Tables~\ref{tab:potentials} and~\ref{tab:potentials1} in Appendix~\ref{app:poten} for all combinations of heavy-antiheavy hadron pairs. 

For a system which can have different $C$-parities, like those in Eqs.~(\ref{eq:cpm}), the potential is expressed as 
\begin{equation}
    V=V_d\pm(-1)^{J-J_1-J_2}c\,c_1c_2V_c
\end{equation}
with \(V_d\) the potential from the {\emph{direct}} process, e.g.,
\(D\bar D^*\to D\bar D^*\) and \(V_c\) from the
{\emph{cross}} one, e.g.,
\(D\bar D^*\to D^*\bar D\). $V_d$'s for these systems are covered by Eq.~({\ref{eq:potential}}), while for the cross processes, it turns out that $V_c$'s for $\ket{D D^*}_c$, $\ket{DD_2}_c$, $\ket{D_1D_2}_c$ and $ \ket{B_6B_{ 6}^{*}}_c$ systems vanish at threshold.  Explicit calculation shows that for the other three systems, $\ket{D D_1}_c$, $\ket{D^*D_1}_c$, and $\ket{D^*D_2}_c$,
\begin{align}
V_c\approx F_c F \zeta_1^2g_V^2 \frac{m_1m_2}{m_{\rm ex}^2-\Delta m^2},\label{eq:Vc}
\end{align}
where $F$ is the same as in Eq.~(\ref{eq:potential}), $\Delta m^2=(m_1-m_2)^2$, and the additional factor $F_c$ accounts for the spin information, which are shown in Appendix~\ref{app:cross}. However, $V_c$ is much smaller than $V_d$ and has little influence on the pole positions compared with the cutoff dependence (see below).

\begin{table}[tb]
    \caption{Values of the coupling parameters used in the calculations.} 
    \centering
    \begin{ruledtabular}
      \begin{tabular}{cccccc}
       $g_V$ &  $\beta$ & $\beta_2$ & $\zeta_1$ & $\beta_B$ & $\beta_S$ \\\hline
       5.8 & 0.9 & $-0.9$ & 0.16 & 0.87 & $-1.74$ \\
       \cite{Bando:1987br} & \cite{Isola:2003fh} & \cite{Dong:2019ofp} & \cite{Dong:2019ofp} & \cite{Liu:2011xc,Chen:2019asm} & \cite{Liu:2011xc,Chen:2019asm}
      \end{tabular}
    \end{ruledtabular}
    \label{tab:parameters}
\end{table}
In Table~\ref{tab:parameters}, we collect the numerical values of the coupling parameters used in our calculations.

\subsection{Potentials from vector charmonia exchange}
\label{sec:ccbarex}

In principle, the $J/\psi$, as well as the excited $\psi$ vector charmonia, can also be exchanged between charmed and anti-charmed hadrons. 
Being vector mesons, their couplings to the charmed mesons have the same spin-momentum behavior as that of the light vectors.
According to Eq.~(\ref{eq:potential}), such contributions should be suppressed  due to the much larger masses of the $\psi$ states than those of the light vectors by a factor of $ {m_\rho^2}/{m_{\psi}^2}\sim 0.1$, up to the difference of coupling constants. Therefore, the exchange of light mesons, if not vanishing, dominates the potentials at threshold. While for the systems where contributions from light vectors vanish (or the $\rho$ and $\omega$ exchanges cancel each other), the vector charmonia exchange, as the sub-leading term, will play an important role in the near-threshold potentials. 

To be more precise, let us take the $J/\psi$ exchange for example, for which the Lagrangian reads
\begin{align}
\mathcal{L}_{{DD} J/\psi}=i g_{DDJ/\psi}\psi^\mu \left(\partial_\mu D^\dagger D-D^\dagger\partial_\mu D\right),
\end{align}
with $g_{DDJ/\psi}\approx 7.64$~\cite{Lin:1999ad}. The resulting potential in the nonrelativistic limit is 
\begin{align}
V\sim -g_{DDJ/\psi}^2\frac{4m_1m_2}{m_{\rm ex}^2},
\end{align}
which is about 40\% of the potential from the $\phi$ exchange between $D_s$ and $\bar D_s$. The contributions from the other vector charmonia will be similar since their masses are of the same order. 
Notice that for all charmed and anti-charmed hadron systems, the vector charmonia exchange yields attractive interactions. Unfortunately, it is not easy to quantitatively estimate their contributions because the masses of these charmonia are much larger than the energy scale 
of interest, and there is no hierarchy among them to help selecting the dominant ones.
Nevertheless, it could be possible to use, e.g., the $Z_c(3900)$, as a benchmark to estimate the overall contribution of the charmonia exchange. 
Given the controversy regarding its pole position~\cite{Ablikim:2013mio,Ablikim:2013xfr,Albaladejo:2015lob,Pilloni:2016obd,Gong:2016jzb}, we refrain from doing so here (for further discussion, see Section~\ref{sec:DD}).

\section{Molecular states from constant interactions}
\label{sec:4}

\subsection{Poles}

Now that we have obtained the constant interactions between a pair of heavy-antiheavy hadrons, we can give a rough picture of the spectrum of possible molecular states. We search for poles of the scattering amplitude by solving the single channel Bethe-Salpeter equation which factorizes into an algebraic equation for a constant potential,
\begin{align}
    T=\frac{V}{1-VG},
\end{align}
where $G$ is the one loop two-body propagator. Here we adopt the dimensional regularization (DR) to regularize the loop integral~\cite{Veltman:1994wz},
\begin{align}
    G(E)=&\frac1{16\pi^2}\bigg\{a(\mu)+\log\frac{m_{1}^2}{\mu^2}+\frac{m_{2}^2-m_{1}^2+s}{2s} \log\frac{m_{2}^2}{m_{1}^2} \nonumber\\
&+\frac{k}{E} \Big[ 
\log\left(2k E+s+\Delta\right) + 
\log\left(2k E+s-\Delta\right) \nonumber\\ 
& -  
\log\left(2k E-s+\Delta\right) - 
\log\left(2k E-s-\Delta\right)
\Big]\bigg\},\label{eq:GDR}
\end{align}
where $s=E^2$, $m_{1}$ and $m_{2}$ are the particle masses, and
\begin{align}
k=\frac1{2E}\lambda^{1/2}(E^2,m_{1}^2,m_{2}^2)\label{eq:3-momentum}
\end{align}
is the corresponding three-momentum with $\lambda(x,y,z)=x^2+y^2+z^2 - 2xy - 2yz - 2xz$ for the K\"all\'en triangle function. Here $\mu$, chosen to be 1 GeV, denotes the DR scale, and $a(\mu)$ is a subtraction constant. The branch cut of $k$ from the threshold to infinity along the positive real $E$ axis splits the whole complex energy plane into two Riemann sheets (RSs) defined as Im$(k)>0$ on the first RS while Im$(k)<0$ on the second RS. Another way to regularize the loop integral is inserting a Gaussian form factor, namely,  
\begin{align}
 G(E) =&\, \int \frac{l^2 dl}{4\pi^2} \frac{\omega_1+\omega_2}{\omega_1\omega_2} 
 \frac{e^{-2l^2/\Lambda^2} }{E^2-(\omega_1+\omega_2)^2+i\epsilon} ,\label{eq:GGF}
\end{align}
with $\omega_i=\sqrt{m_i^2+l^2}$.  The cutoff $\Lambda$ is usually in the range of $0.5\sim 1.0$ GeV. The subtraction constant $a(\mu)$ in DR is determined by matching the values of $G$ from these two methods at threshold. We will use the DR loop with the so-determined subtraction constant for numerical calculations.

\begin{figure*}[tbhp]
    \centering
    \includegraphics[width=0.9\linewidth]{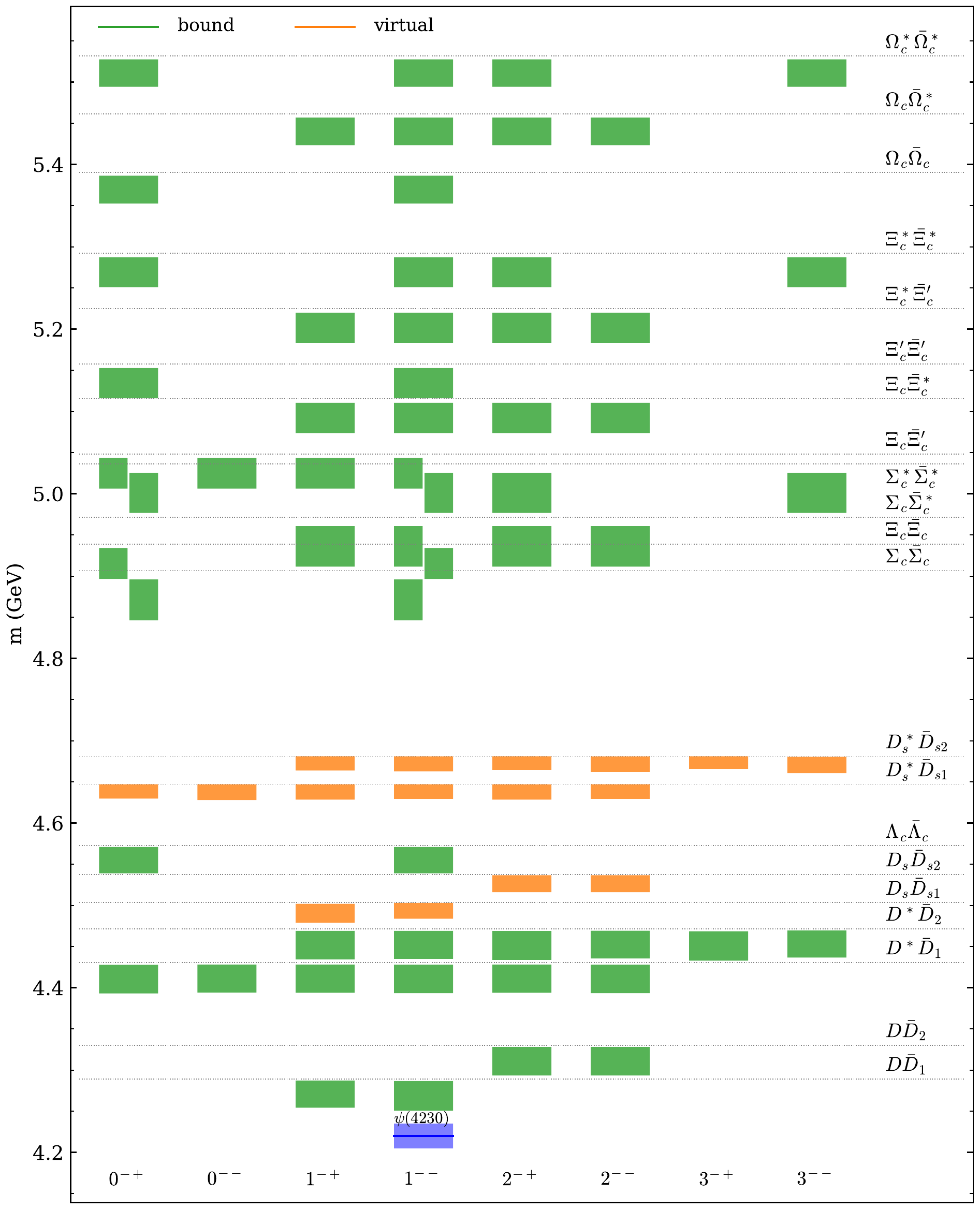}
    \caption{The spectrum of hadronic molecules consisting of a pair of charmed-anticharmed hadrons with $(I,S)=(0,0)$.
    $0^{--}$, $1^{-+}$ and $3^{-+}$ are exotic quantum numbers.
    The colored rectangle, green for a bound state and orange for a virtual state, covers the range of the pole position for a given system with cutoff $\Lambda$ varies in the range of $[0.5, 1.0]$~GeV. Thresholds are marked by dotted horizontal lines.
    The rectangle closest to, but below, the threshold corresponds to the hadronic molecule in that system. In some cases where the pole positions of two systems overlap, small rectangles are used with the left (right) one for the system with the higher (lower) threshold. The blue line (band) represents the center value (error) of the mass of the experimental candidate of the corresponding molecule. The averaged central value and error of the $\psi(4230)$ mass are taken from RPP~\cite{Zyla:2020zbs}.}\label{fig:specBB00}
\end{figure*}

\begin{figure*}
    \centering
    \includegraphics[width=0.85\linewidth]{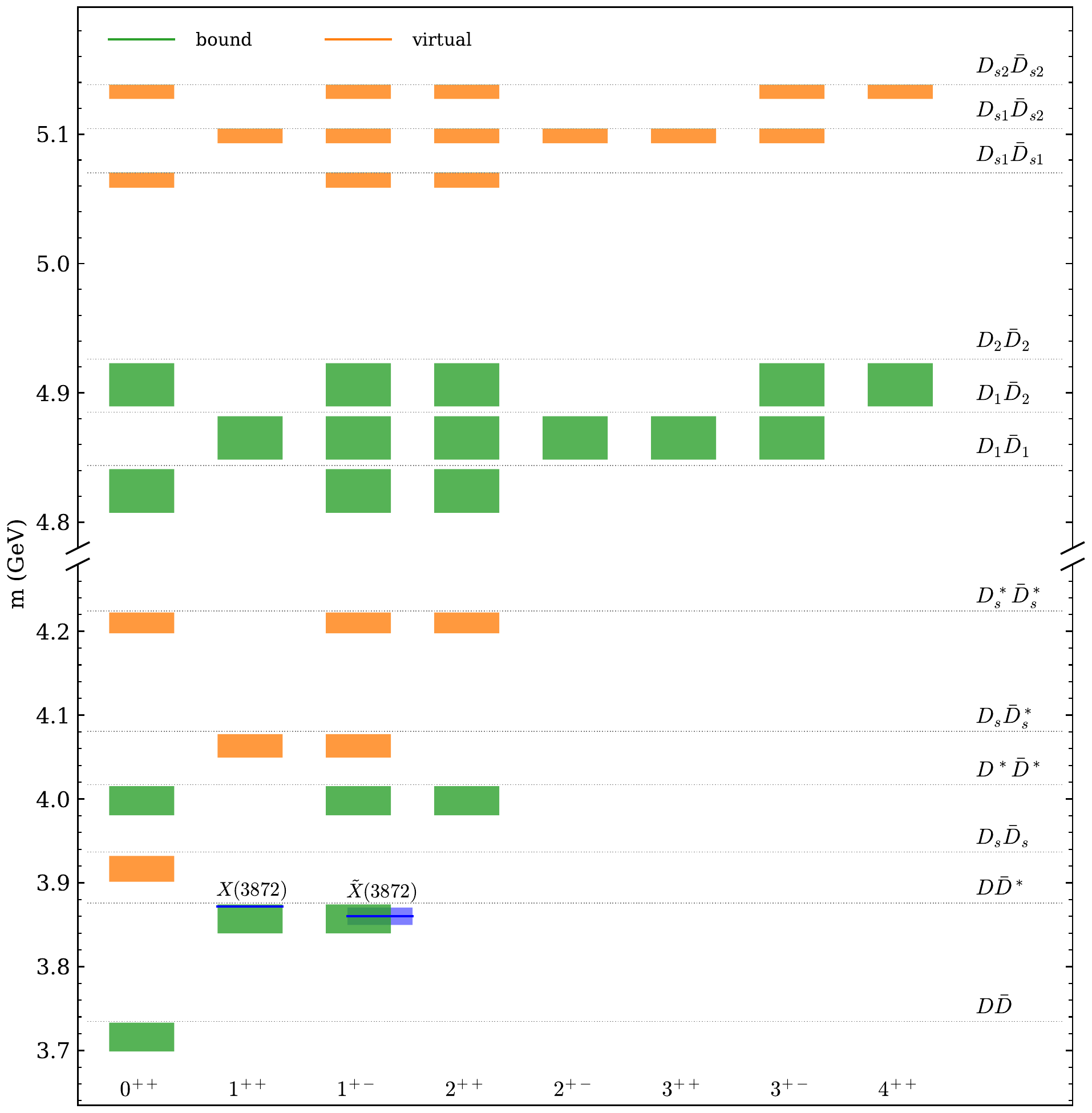}
    \caption{The spectrum of hadronic molecules consisting of a pair of charmed-anticharmed hadrons with $(I,S)=(0,0)$. $2^{+-}$ are exotic quantum numbers. The parameters of the $X(3872)$ and $\tilde X(3872)$ are taken from  RPP~\cite{Zyla:2020zbs} and Ref.~\cite{Aghasyan:2017utv}, respectively. See the caption for Fig.\ref{fig:specBB00}. }\label{fig:specDDD1D100}
\end{figure*}

\begin{figure*}[tbhp]
        \centering
        \includegraphics[height=6.7cm]{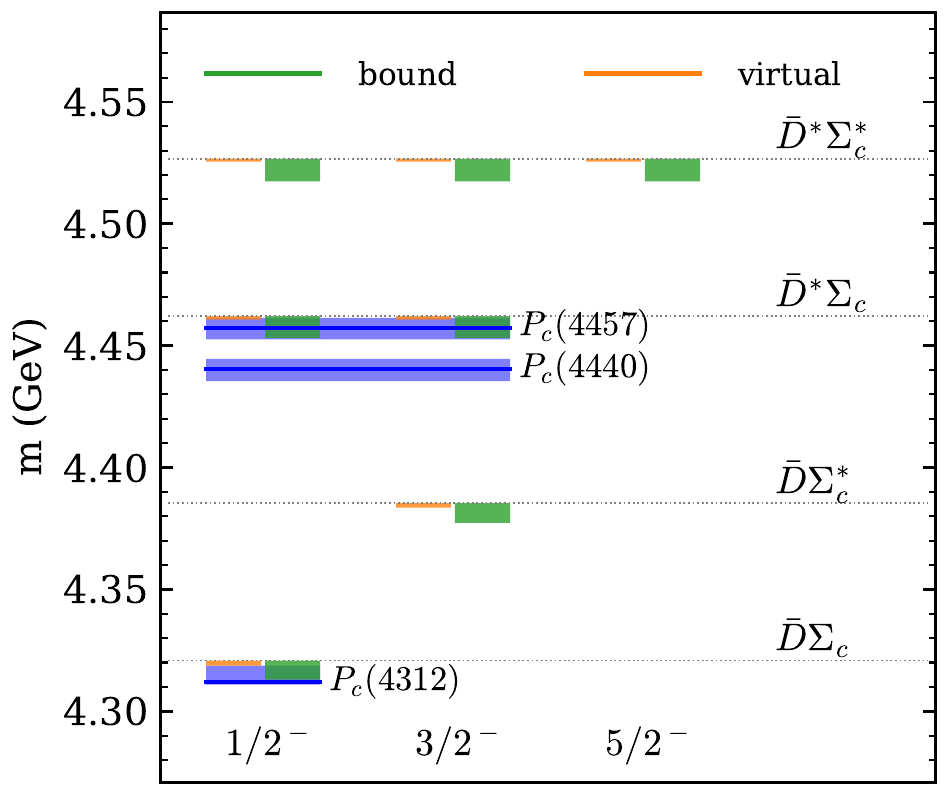}
        \includegraphics[height=6.7cm]{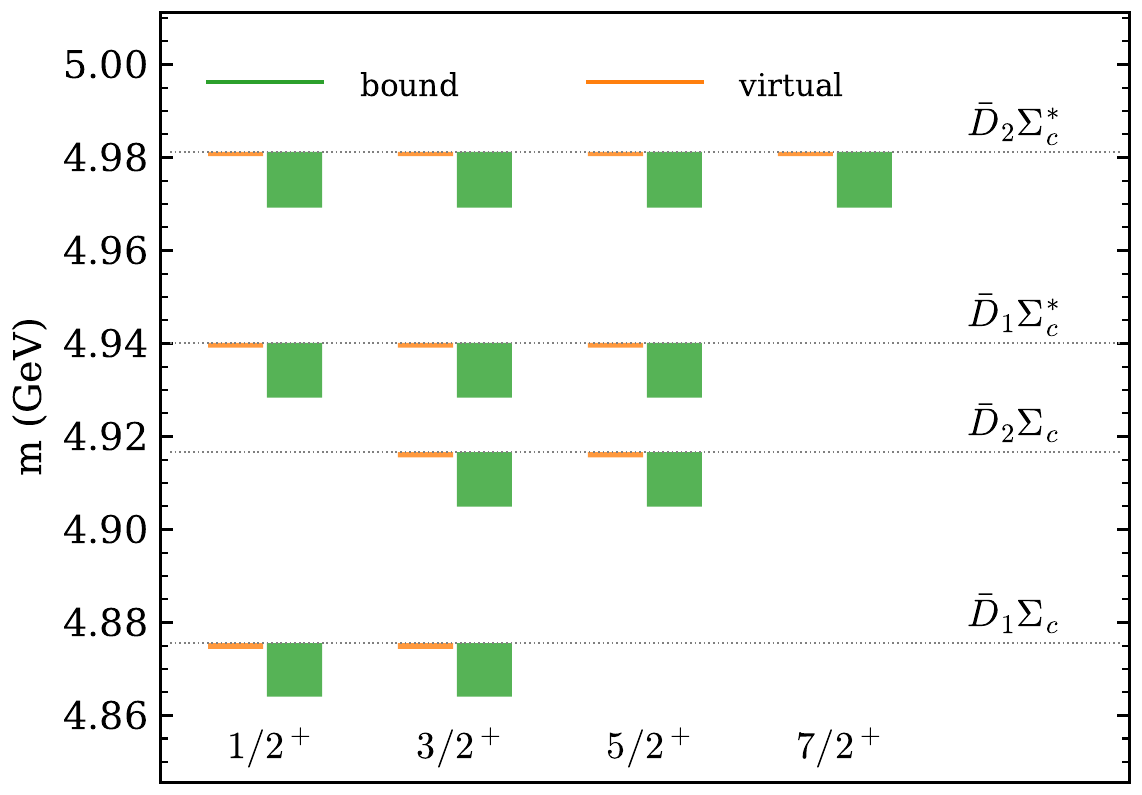}
        \caption{The spectrum of hadronic molecules consisting of a pair of charmed-anticharmed hadrons with $(I,S)=(\frac12,0)$ and unit baryon number. 
        The left orange band and right green band for each pole represent that the pole moves from a virtual state on the second RS to a bound state on the first RS when the cutoff $\Lambda$ changes from 0.5 to 1.0~GeV. The parameters of these three $P_{c}$ states are taken from Ref.~\cite{Aaij:2019vzc}, whose $J^P$ have not been determined experimentally. See the caption for Fig.~\ref{fig:specBB00}.} \label{fig:specDB50}
\end{figure*}

\begin{figure*}[tbhp]
    \centering
    \includegraphics[height=6.7cm]{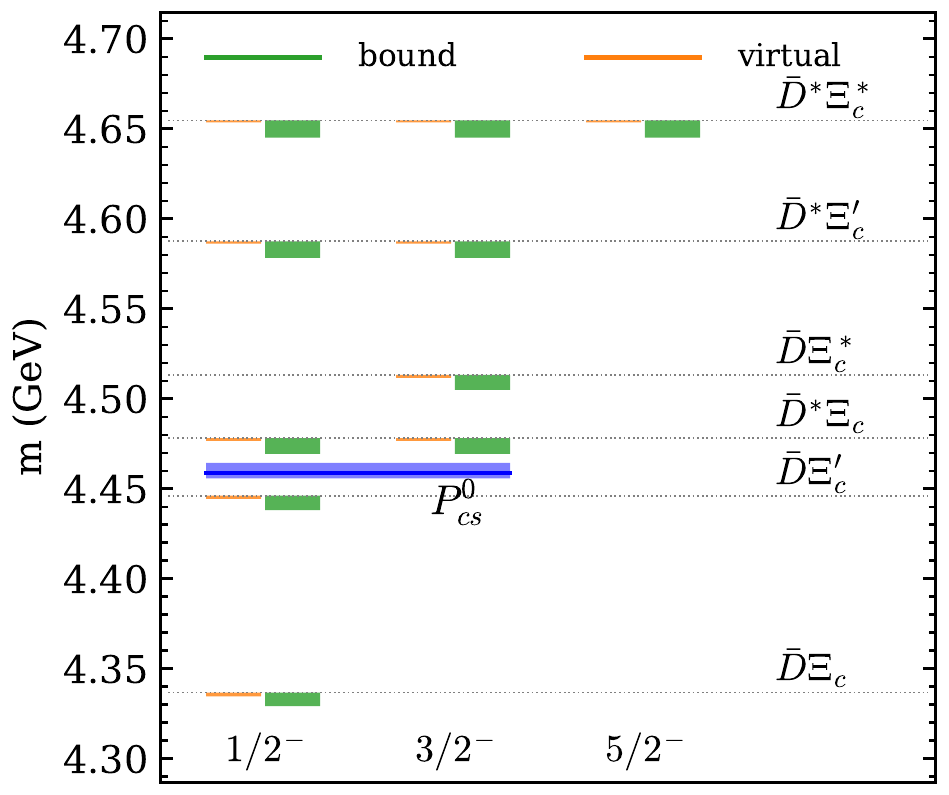}
    \includegraphics[height=6.7cm]{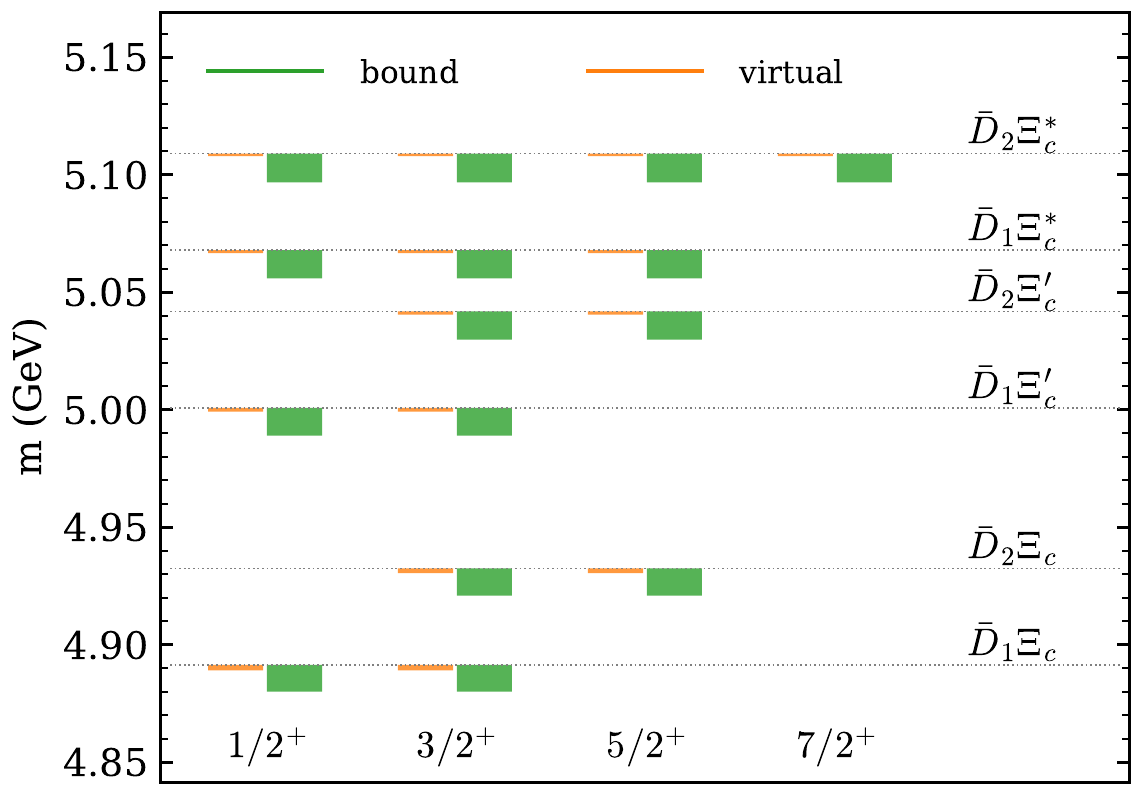}
    \caption{The spectrum of hadronic molecules consisting of a pair of charmed-anticharmed hadrons with $(I,S)=(0,1)$ and unit baryon number.  
    The parameters of $P_{cs}(4459)$ are taken from Ref.~\cite{Aaij:2020gdg}, whose $J^P$ have not been determined experimentally. See the caption for Fig.~\ref{fig:specDB50}.}\label{fig:specDB01}
\end{figure*}

\begin{figure*}[tbhp]
    \centering
    \includegraphics[width=0.9\linewidth]{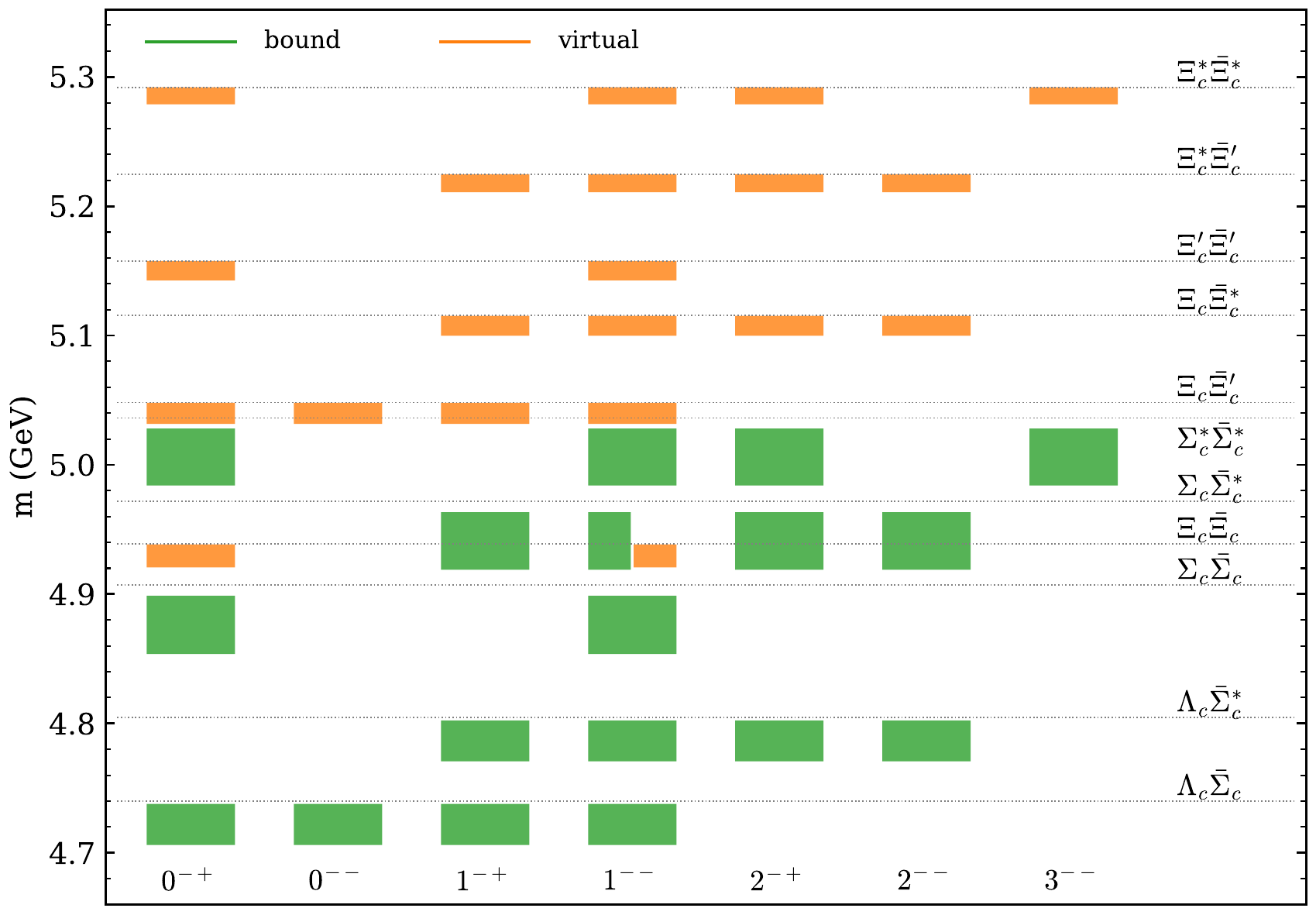}
    \caption{The spectrum of hadronic molecules consisting of a pair of charmed-anticharmed hadrons with $(I,S)=(1,0)$. See the caption for Fig.~\ref{fig:specBB00}.}\label{fig:BB10}
\end{figure*}

\begin{figure*}[tbhp]
    \centering
    \includegraphics[height=8.1cm]{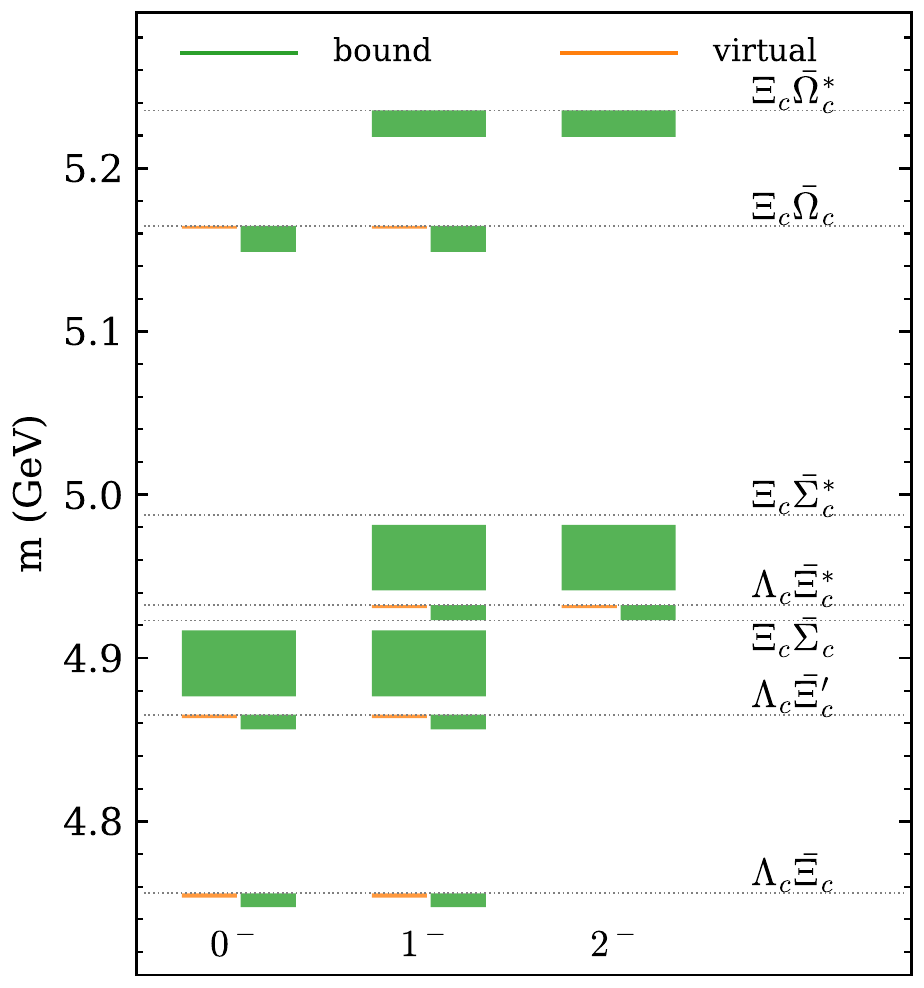}
    \includegraphics[height=8.1cm]{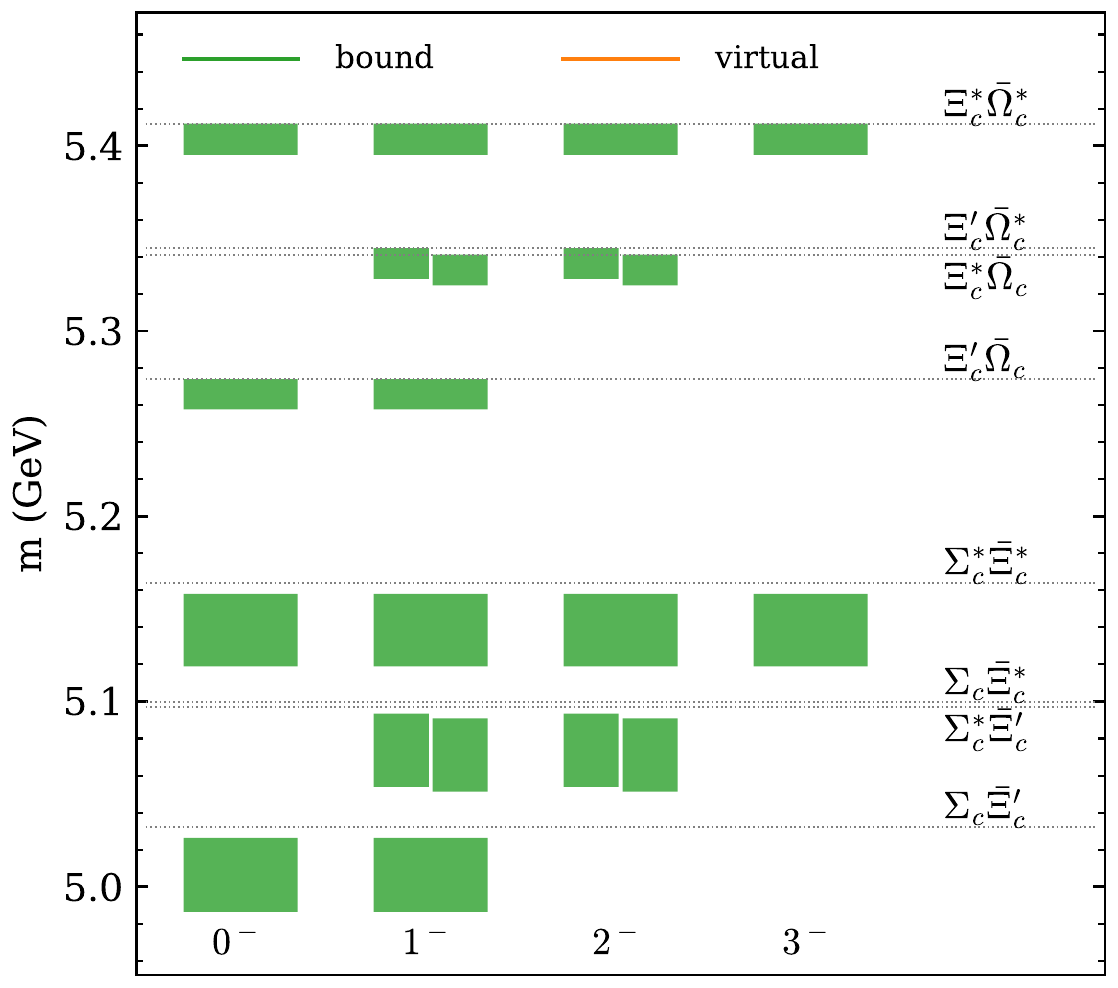}
    \caption{The spectrum of hadronic molecules consisting of a pair of charmed-anticharmed hadrons with $(I,S)=(\frac12,1)$. See the captions for Figs.~\ref{fig:specBB00} and \ref{fig:specDB50}.}\label{fig:TS00}
\end{figure*}

For a single channel, if the interaction is attractive and strong enough to form a bound state, the pole will be located below threshold on the first RS. If it is not strong enough, the pole will move onto the second RS as a virtual state, still below threshold. 
In Tables~\ref{tab:pole00}, \ref{tab:pole00c}, \ref{tab:pole50}, \ref{tab:pole01},  \ref{tab:pole51} and \ref{tab:pole10}, we list all the pole positions of the heavy-antiheavy hadron systems which have attractive interactions, corresponding to the masses of hadronic molecules. 
For better illustration, these states, together with some hadronic molecule candidates observed in experiments, are also shown in Figs.~\ref{fig:specBB00}, \ref{fig:specDDD1D100}, \ref{fig:specDB50}, \ref{fig:specDB01}, \ref{fig:BB10} and \ref{fig:TS00}. 
In total, we obtain a spectrum of 229  hadronic molecules
considering constant contact interactions, saturated by the light vector mesons, with the coupled-channel effects neglected. 

\begin{table}[tb]
\caption{Pole positions of heavy-antiheavy hadron systems with $(I,S)=(0,0)$. $E_{\rm th}$ in the second column is the threshold in MeV. The number 0.5 (1.0) in the third (fourth) column means that the cutoff $\Lambda=0.5$ ($1.0$)~GeV for Eq.~(\ref{eq:GGF}) is used to determine the subtraction constant $a(\mu)$ in Eq.~(\ref{eq:GDR}). In the last two columns, the first number in the parenthesis refers to the RS where the pole is located while the second number means the distance between the pole position and the corresponding threshold, namely, $E_{\rm th}-E_{\rm pole}$, in MeV.} \label{tab:pole00}
\centering
\begin{ruledtabular}
\begin{tabular}{l|cccc}
System & $E_{\rm th}$ & $J^{PC}$&Pole (0.5) & Pole (1.0)\\
\hline
$D\bar D$&3734&$0^{++}$&(1, 1.31)&(1, 35.8)\\
$D\bar D^*$&3876&$1^{+\pm}$&(1, 1.56)&(1, 36.2)\\
$D_s\bar D_s$&3937&$0^{++}$&(2, 35.5)&(2, 4.72)\\
$D^*\bar D^*$&4017&$(0,2)^{++},1^{+-}$&(1, 1.82)&(1, 36.6)\\
$D_s\bar D_s^*$&4081&$1^{+\pm}$&(2, 31.0)&(2, 3.15)\\
$D_s^*\bar D_s^*$&4224&$(0,2)^{++},1^{+-}$&(2, 26.7)&(2, 1.92)\\
$D\bar D_2$&4330&$2^{-\pm}$&(1, 2.2)&(1, 36.7)\\
$D_s\bar D_{s2}$&4537&$2^{-\pm}$&(2, 21.3)&(2, 0.713)\\
$D_1\bar D_1$&4844&$(0,2)^{++},1^{+-}$&(1, 3.01)&(1, 36.7)\\
$D_1\bar D_2$&4885&$(1,2,3)^{+\pm}$&(1, 3.06)&(1, 36.6)\\
$D_2\bar D_2$&4926&$(0,2,4)^{++},(1,3)^{+-}$&(1, 3.1)&(1, 36.6)\\
$D_{s1}\bar D_{s1}$&5070&$(0,2)^{++},1^{+-}$&(2, 11.7)&(1, 0.074)\\
$D_{s1}\bar D_{s2}$&5104&$(1,2,3)^{+\pm}$&(2, 11.3)&(1, 0.104)\\
$D_{s2}\bar D_{s2}$&5138&$(0,2,4)^{++},(1,3)^{+-}$&(2, 10.9)&(1, 0.139)\\
\hline
$\Lambda_c \bar \Lambda_c$&4573&$0^{-+},1^{--}$&(1, 1.98)&(1, 33.8)\\
$\Sigma_c \bar \Sigma_c$&4907&$0^{-+},1^{--}$&(1, 11.1)&(1, 60.8)\\
$\Xi_c \bar \Xi_c$&4939&$0^{-+},1^{--}$&(1, 4.72)&(1, 42.2)\\
$\Sigma_c^* \bar \Sigma_c$&4972&$1^{-\pm},2^{-\pm}$&(1, 11.0)&(1, 60.1)\\
$\Sigma_c^* \bar \Sigma_c^*$&5036&$(0,2)^{-+},(1,3)^{--}$&(1, 10.9)&(1, 59.5)\\
$\Xi_c \bar \Xi_c'$&5048&$0^{-\pm},1^{-\pm}$&(1, 4.79)&(1, 41.9)\\
$\Xi_c \bar \Xi_c^*$&5115&$1^{-\pm},2^{-\pm}$&(1, 4.84)&(1, 41.6)\\
$\Xi_c' \bar \Xi_c'$&5158&$0^{-+},1^{--}$&(1, 4.87)&(1, 41.5)\\
$\Xi_c^* \bar \Xi_c'$&5225&$1^{-\pm},2^{-\pm}$&(1, 4.91)&(1, 41.3)\\
$\Xi_c^* \bar \Xi_c^*$&5292&$(0,2)^{-+},(1,3)^{--}$&(1, 4.95)&(1, 41.0)\\
$\Omega_c \bar \Omega_c$&5390&$0^{-+},1^{--}$&(1, 4.17)&(1, 38.0)\\
$\Omega_c^* \bar \Omega_c$&5461&$1^{-\pm},2^{-\pm}$&(1, 4.22)&(1, 37.8)\\
$\Omega_c^* \bar \Omega_c^*$ &5532&$(0,2)^{-+},(1,3)^{--}$&(1, 4.26)&(1, 37.6)\\
\end{tabular}
\end{ruledtabular}
\end{table}

\begin{table}[tbh]
\caption{Pole positions of heavy-antiheavy hadron systems with $(I,S)=(0,0)$. See the caption for Table \ref{tab:pole00}. In these systems different total spins and $C$-parities yield slightly different pole positions.} \label{tab:pole00c}
\centering
\begin{ruledtabular}
\begin{tabular}{l|cccc}
System & $E_{\rm th}$ & $J^{PC}$&Pole (0.5) & Pole (1.0)\\
\hline
$D\bar D_1$&4289&$1^{-+}$&(1, 1.78)&(1, 34.9)\\
& & $1^{--}$&(1, 2.53)&(1, 38.4)\\
\hline
$D^*\bar D_1$&4431&$0^{-+}$&(1, 2.55)&(1, 37.4)\\
& & $0^{--}$&(1, 2.29)&(1, 36.3)\\
&&$1^{-+}$&(1, 2.36)&(1, 36.6)\\
& & $1^{--}$&(1, 2.49)&(1, 37.1)\\
&&$2^{-+}$&(1, 2.36)&(1, 36.6)\\
& & $2^{--}$&(1, 2.49)&(1, 37.1)\\
\hline
$D^*\bar D_2$&4472&$1^{-+}$&(1, 2.54)&(1, 37.1)\\
& & $1^{--}$&(1, 2.4)&(1, 36.5)\\
&&$2^{-+}$&(1, 2.68)&(1, 37.7)\\
& & $2^{--}$&(1, 2.26)&(1, 35.9)\\
&&$3^{-+}$&(1, 2.89)&(1, 38.6)\\
& & $3^{--}$&(1, 2.05)&(1, 34.9)\\
\hline
$D_s \bar D_{s1}$&4503&$1^{-+}$&(2, 24.2)&(2, 1.4)\\
& & $1^{--}$&(2, 19.6)&(2, 0.402)\\
\hline
$D_s^*\bar D_{s1}$&4647&$0^{-+}$&(2, 17.5)&(2, 0.179)\\
& & $0^{--}$&(2, 19.2)&(2, 0.402)\\
&&$1^{-+}$&(2, 18.7)&(2, 0.402)\\
& & $1^{--}$&(2, 17.9)&(2, 0.227)\\
&&$2^{-+}$&(2, 18.7)&(2, 0.342)\\
& & $2^{--}$&(2, 17.9)&(2, 0.402)\\
\hline
$D_s^*\bar D_{s2}$&4681&$1^{-+}$&(2, 17.4)&(2, 0.177)\\
& & $1^{--}$&(2, 18.3)&(2, 0.402)\\
&&$2^{-+}$&(2, 16.6)&(2, 0.402)\\
& & $2^{--}$&(2, 19.2)&(2, 0.418)\\
&&$3^{-+}$&(2, 15.4)&(2, 0.023)\\
& & $3^{--}$&(2, 20.6)&(2, 0.402)\\
\end{tabular}
\end{ruledtabular}
\end{table}

\begin{table}[th]
    \caption{Pole positions of heavy-antiheavy hadron systems with $(I,S)=(1/2,0)$ and unit baryon number. See the caption for Table~\ref{tab:pole00}.} \label{tab:pole50}
    \centering
    \begin{ruledtabular}
    \begin{tabular}{l|cccc}
    System & $E_{\rm th}$ &$J^P$ & Pole (0.5) & Pole (1.0)\\
    \hline
    $\bar D\Sigma_c$&4321&$\frac{1}{2}^{-}$&(2, 2.04)&(1, 7.79)\\
    $\bar D\Sigma_c^*$&4385&$\frac{3}{2}^{-}$&(2, 1.84)&(1, 8.1)\\
    $\bar D^*\Sigma_c$&4462&$(\frac{1}{2},\frac{3}{2})^{-}$&(2, 1.39)&(1, 8.95)\\
    $\bar D^*\Sigma_c^*$&4527&$(\frac{1}{2},\frac{3}{2},\frac{5}{2})^{-}$&(2, 1.23)&(1, 9.26)\\
    $\bar D_1\Sigma_c$&4876&$(\frac{1}{2},\frac{3}{2})^{+}$&(2, 0.417)&(1, 11.5)\\
    $\bar D_2\Sigma_c$&4917&$(\frac{3}{2},\frac{5}{2})^{+}$&(2, 0.366)&(1, 11.7)\\
    $\bar D_1\Sigma_c^*$&4940&$(\frac{1}{2},\frac{3}{2},\frac{5}{2})^{+}$&(2, 0.34)&(1, 11.8)\\
    $\bar D_2\Sigma_c^*$&4981&$(\frac{1}{2},\frac{3}{2},\frac{5}{2},\frac{7}{2})^{+}$&(2, 0.294)&(1, 12.0)\\
    \end{tabular}
    \end{ruledtabular}
 \end{table}

\begin{table}[h]
\caption{Pole positions of heavy-antiheavy hadron systems with $(I,S)=(0,1)$ and unit baryon number. See the caption for Table~\ref{tab:pole00}.} \label{tab:pole01}
\centering
\begin{ruledtabular}
\begin{tabular}{l|cccc}
System & $E_{\rm th}$& $J^P$ & Pole (0.5) & Pole (1.0)\\
\hline
$\bar D \Xi_c$&4337&$\frac{1}{2}^{-}$&(2, 2.14)&(1, 7.53)\\
$\bar D  \Xi_c'$&4446&$\frac{1}{2}^{-}$&(2, 1.82)&(1, 8.05)\\
$\bar D^* \Xi_c$&4478&$(\frac{1}{2},\frac{3}{2})^{-}$&(2, 1.47)&(1, 8.69)\\
$\bar D \Xi_c^*$&4513&$\frac{3}{2}^{-}$&(2, 1.65)&(1, 8.34)\\
$\bar D^* \Xi_c'$&4587&$(\frac{1}{2},\frac{3}{2})^{-}$&(2, 1.21)&(1, 9.21)\\
$\bar D^* \Xi_c^*$&4655&$(\frac{1}{2},\frac{3}{2},\frac{5}{2})^{-}$&(2, 1.08)&(1, 9.51)\\
$\bar D_1\Xi_c$&4891&$(\frac{1}{2},\frac{3}{2})^{+}$&(2, 0.455)&(1, 11.3)\\
$\bar D_2 \Xi_c$&4932&$(\frac{3}{2},\frac{5}{2})^{+}$&(2, 0.4)&(1, 11.5)\\
$\bar D_1 \Xi_c'$&5001&$(\frac{1}{2},\frac{3}{2})^{+}$&(2, 0.326)&(1, 11.8)\\
$\bar D_2 \Xi_c'$&5042&$(\frac{3}{2},\frac{5}{2})^{+}$&(2, 0.28)&(1, 12.0)\\
$\bar D_1 \Xi_c^*$&5068&$(\frac{1}{2},\frac{3}{2},\frac{5}{2})^{+}$&(2, 0.262)&(1, 12.1)\\
$\bar D_2 \Xi_c^*$&5109&$(\frac{1}{2},\frac{3}{2},\frac{5}{2},\frac{7}{2})^{+}$&(2, 0.222)&(1, 12.3)\\
\end{tabular}
\end{ruledtabular}
\end{table}

\begin{table}[th]
\caption{Pole positions of heavy-antiheavy hadron systems with $(I,S)=(1/2,1)$. See the caption for Table \ref{tab:pole00}.} \label{tab:pole51}
\centering
\begin{ruledtabular}
\begin{tabular}{l|cccc}
System & $E_{\rm th}$ &$J^P$ & Pole (0.5) & Pole (1.0)\\
\hline
$\Lambda_c\bar\Xi_c$&4756&$(0,1)^-$&(2, 1.29)&(1, 8.42)\\
$\Lambda_c\bar\Xi_c'$&4865&$(0,1)^-$&(2, 1.05)&(1, 8.93)\\
$\Xi_c\bar\Sigma_c$&4923&$(0,1)^-$&(1, 5.98)&(1, 46.4)\\
$\Lambda_c\bar\Xi_c^*$&4932&$(1,2)^-$&(2, 0.92)&(1, 9.23)\\
$\Xi_c\bar\Sigma_c^*$&4988&$(1,2)^-$&(1, 6.01)&(1, 46.1)\\
$\Sigma_c\bar\Xi_c'$&5032&$(0,1)^-$&(1, 6.03)&(1, 45.9)\\
$\Sigma_c^*\bar\Xi_c'$&5097&$(1,2)^-$&(1, 6.06)&(1, 45.6)\\
$\Sigma_c\bar\Xi_c^*$&5100&$(1,2)^-$&(1, 6.05)&(1, 45.6)\\
$\Sigma_c^*\bar\Xi_c^*$&5164&$(0,1,2,3)^-$&(1, 6.08)&(1, 45.2)\\
$\Xi_c\bar\Omega_c$&5165&$(0,1)^-$&(2, 8e-5)&(1, 15.9)\\
$\Xi_c\bar\Omega_c^*$&5235&$(1,2)^-$&(1, 0.002)&(1, 16.2)\\
$\Xi_c'\bar\Omega_c$&5274&$(0,1)^-$&(1, 0.006)&(1, 16.4)\\
$\Xi_c^*\bar\Omega_c$&5341&$(1,2)^-$&(1, 0.016)&(1, 16.6)\\
$\Xi_c'\bar\Omega_c^*$&5345&$(1,2)^-$&(1, 0.016)&(1, 16.6)\\
$\Xi_c^*\bar\Omega_c^*$&5412&$(0,1,2,3)^-$&(1, 0.030)&(1, 16.8)\\
\end{tabular}
\end{ruledtabular}
\end{table}

\ 
\begin{table}[tbhp]
\caption{Pole positions of heavy-antiheavy hadron systems with $(I,S)=(1,0)$. See the caption for Table \ref{tab:pole00}.} \label{tab:pole10}
\centering
\begin{ruledtabular}
\begin{tabular}{l|cccc}
System & $E_{\rm th}$ &$J^{PC}$  & Pole (0.5) & Pole (1.0)\\
\hline
$\Lambda_c\bar \Sigma_c$&4740&$(0,1)^{-\pm}$&(1, 2.19)&(1, 33.9)\\
$\Lambda_c\bar \Sigma_c^*$&4805&$(1,2)^{-\pm}$&(1, 2.27)&(1, 33.9)\\
$\Sigma_c\bar \Sigma_c$&4907&$0^{-+},1^{--}$&(1, 8.28)&(1, 53.3)\\
$\Xi_c\bar \Xi_c$&4939&$0^{-+},1^{--}$&(2, 18.2)&(2, 0.39)\\
$\Sigma_c^*\bar \Sigma_c$&4972&$(1,2)^{-\pm}$&(1, 8.27)&(1, 52.8)\\
$\Sigma_c^*\bar \Sigma_c^*$&5036&$(0,2)^{-+},(1,3)^{--}$&(1, 8.25)&(1, 52.3)\\
$\Xi_c\bar \Xi_c'$&5048&$(0,1)^{-\pm}$&(2, 16.5)&(2, 0.19)\\
$\Xi_c\bar \Xi_c^*$&5115&$(1,2)^{-\pm}$&(2, 15.6)&(2, 0.11)\\
$\Xi_c'\bar \Xi_c'$&5158&$0^{-+},1^{--}$&(2, 14.9)&(2, 0.061)\\
$\Xi_c^*\bar \Xi_c'$&5225&$(1,2)^{-\pm}$&(2, 14.0)&(2, 0.020)\\
$\Xi_c^*\bar \Xi_c^*$&5292&$(0,2)^{-+},(1,3)^{--}$&(2, 13.2)&(2, 0.002)\\

\end{tabular}
\end{ruledtabular}
\end{table}

\subsection{Discussions of selected systems}

It is worthwhile to notice that the overwhelming majority of the predicted spectrum is located in the energy region that has not been experimentally explored in detail. 
Searching for these states at BESIII, Belle-II, LHCb and other planned experiments will be important to establish a clear pattern of the hidden-charm states and to understand how QCD organizes the hadron spectrum.

In the following, we discuss a few interesting systems that have experimental candidates.

\subsubsection{$D^{(*)}\bar D^{(*)}$: $X(3872)$, $Z_c(3900)$ and their partners}
\label{sec:DD}

Within the mechanism considered here, the interactions of $D\bar D$, $D\bar D^*$ and $D^*\bar D^*$ are the same. 
For the meson pairs to be isospin scalars, the attractions are strong enough to form bound states with similar binding energies, see Table~\ref{tab:pole00} and Fig.~\ref{fig:specDDD1D100}, while for the isovector pairs, the contributions from the $\rho$ and $\omega$ exchanges cancel each other, see Table~\ref{tab:potentials} in Appendix~\ref{app:poten}.

The $X(3872)$ observed by the Belle Collaboration~\cite{Choi:2003ue} is widely suggested as an isoscalar $D\bar D^*$ molecule with $J^{PC}=1^{++}$~\cite{Tornqvist:2003na,Wong:2003xk,Swanson:2003tb,Tornqvist:2004qy} (for reviews, see, e.g., Refs.~\cite{Chen:2016qju,Guo:2017jvc,Kalashnikova:2018vkv,Brambilla:2019esw}). 
Actually such a hadronic molecule was predicted 10 years before the discovery by T\"ornvist considering the one-pion exchange~\cite{Tornqvist:1993ng}. 
Our results show that the light vector exchange leads to a near-threshold isoscalar bound state that can be identified with the $X(3872)$ as well, together with a negative $C$-parity partner of $X(3872)$ with the same binding energy (see also Refs.~\cite{Gamermann:2006nm,Gamermann:2007fi}). There is experimental evidence of such a negative $C$-parity state, named as $\tilde X(3872)$,\footnote{It should be called $h_c(3872)$ according to the RPP nomenclature.} reported by the COMPASS Collaboration~\cite{Aghasyan:2017utv}.
A recent study of the $D_{(s)}^{(*)}\bar D_{(s)}^{(*)}$ molecular states using the method of QCD sum rules also finds both $1^{++}$ and $1^{+-}$ $D\bar D^*$ states~\cite{Wang:2020dgr}.

The potential predicting the $X(3872)$ as an isoscalar $D\bar D^*$ bound state, also predicts the existence of isoscalar $D\bar D$ and $D^*\bar D^*$ bound states. 
By imposing only HQSS, there are two independent contact terms in the $D^{(*)}\bar D^{(*)}$ interactions for each isospin~\cite{AlFiky:2005jd,Nieves:2012tt}, which can be defined as~\cite{Guo:2017jvc}
\begin{align}
    C_{H\bar H,0} &= \bra{\frac12,\frac12,0}\mathcal{H}_I\ket{\frac12,\frac12,0}, \notag\\
    C_{H\bar H,1} &= \bra{\frac12,\frac12,1}\mathcal{H}_I\ket{\frac12,\frac12,1},
\end{align}
where $\mathcal{H}_I$ is the interaction Hamiltonian, and $\ket{s_{\ell1},s_{\ell2},s_\ell}$ denotes the charmed meson pair with $s_{\ell}$ being the total angular momentum of the light degrees of freedom in the two-meson system and $s_{\ell1,\ell2}$ for the individual mesons.
Such an analysis leads to the prediction of a $2^{++}$ $D^*\bar D^*$ tensor state as the HQSS partner of the $X(3872)$ considering the physical charmed meson masses~\cite{Nieves:2012tt,HidalgoDuque:2012pq,Guo:2013sya} and three partners with $0^{++}, 1^{+-}$ and $2^{++}$ in the strict heavy quark limit~\cite{Hidalgo-Duque:2013pva,Baru:2016iwj} that depend on the same contact term as the $X(3872)$.
The resonance saturation by the light vector mesons in fact leads to a relation $C_{H\bar H,0} = C_{H\bar H,1}$, and consequently 6 $S$-wave $D^{(*)}\bar D^{(*)}$ bound states.

The existence of an isoscalar $D\bar D$ bound state has been predicted by various phenomenology models~\cite{Zhang:2006ix,Gamermann:2006nm,Liu:2008tn,Wong:2003xk,Nieves:2012tt,HidalgoDuque:2012pq}, and more recently by lattice QCD calculations~\cite{Prelovsek:2020eiw}. 
Despite attempts~\cite{Gamermann:2007mu,Dai:2020yfu,Wang:2020elp} to dig out hints for such a state from the available experimental data~\cite{Uehara:2005qd,Abe:2007sya,Aubert:2010ab}, no clear evidence has yet been found. However, this could be because its mass is below the $D\bar D$ threshold so that no easily detectable decay modes are available.

As for the isoscalar $2^{++}$ $D^*\bar D^*$ bound state, it can decay into $D\bar D$ in $D-$wave, and the width was predicted to be in the range from a few to dozens of MeV~\cite{Albaladejo:2015dsa,Baru:2016iwj}. No evidence has been found so far. One possible reason is that the coupling to ordinary charmonia could either move the $2^{++}$ pole deep into the complex energy plane and thus invisible~\cite{Cincioglu:2016fkm}\footnote{For a discussion of the intricate interplay between a meson-meson channel with multiple quark model states, see Ref.~\cite{Hammer:2016prh}.} or make the $D^*\bar D^*$ interaction in the $2^{++}$ sector unbound~\cite{Ortega:2017qmg,Ortega:2020tng}.\footnote{Mixing of two energy levels will push them further apart. Thus,  mixing of the $D^*\bar D^*$ state with a lower-mass $\chi_{c2}(2P)$ can effectively provide a repulsive contribution to the $D^*\bar D^*$ interaction.} 
For more discussions regarding the mixing of charmonia with meson-meson channels, we refer to Refs.~\cite{Kalashnikova:2005ui,Zhou:2017dwj,Cincioglu:2020amz}.

The $Z_c(3900)$~\cite{Ablikim:2013mio,Liu:2013dau,Ablikim:2013xfr} was also suggested to be an isovector $D\bar D^*$ molecule with quantum numbers $J^{PC}=1^{+-}$~\cite{Wang:2013cya,Guo:2013sya} even though the light vector exchange vanishes in this case. Recall that the vector charmonia exchange will also yield an attractive interaction, as discussed in Section~\ref{sec:ccbarex}, which can possibly lead to a virtual state below threshold. In fact, it has been suggested that  the $J/\psi$-exchange is essential in the formation of the $Z_c(3900)$ in Ref.~\cite{Aceti:2014uea}.
It was shown in Refs.~\cite{Albaladejo:2015lob, He:2017lhy,Ortega:2018cnm} that a virtual state assignment for the $Z_c(3900)$ is consistent with the experimental data.\footnote{Notice that the coupling of the $D\bar D^*$ to a lower channel, which is $J/\psi\pi$ in this case, induces a finite width for the virtual state pole so that it behaves like a resonance, i.e. a pole in the complex plane off the real axis. The same is true for all other poles generated here.} Furthermore, it was shown in Ref.~\cite{Albaladejo:2016jsg} that the finite volume energy levels are also consistent with the lattice QCD results which did not report an additional state~\cite{Prelovsek:2014swa}. 
Similarly, the $Z_c(4020)^\pm$~\cite{Ablikim:2013wzq, Ablikim:2013emm} with isospin-1 near the $D^*\bar D^*$ threshold can be a virtual state as well.
It was recently argued that a near-threshold virtual state needs to be understood as a hadronic molecule~\cite{Matuschek:2020gqe}.
Analysis of the Belle data on the $Z_b$ states~\cite{Belle:2011aa,Garmash:2015rfd} using constant contact terms to construct the unitary $T$-matrix also supports the $Z_b$ states as hadronic molecules~\cite{Cleven:2011gp,Hanhart:2015cua,Guo:2016bjq,Wang:2018jlv,Baru:2020ywb}.
The molecular explanation of the $Z_c$ and $Z_b$ states is further supported by their decay patterns studied using a quark exchange model~\cite{Wang:2018pwi,Xiao:2019spy}.
Without a quantitative calculation, as commented in Section~\ref{sec:ccbarex}, we postulate that
there can be 6 isovector hadronic molecules (with the same $J^{PC}$ as the isoscalar ones) as virtual states of $D^{(*)}\bar D^{(*)}$, which will show up as prominent threshold cusps (see Ref.~\cite{Dong:2020hxe} for a general discussion of the line shape behavior in the near-threshold region).

\subsubsection{$D_{s}^{(*)}\bar D_s^{(*)}$ virtual states} 
\label{sec:dsds}

Here we find that the potential from the $\phi$ exchange is probably not enough to form bound states of $D_s^{(*)}\bar D_s^{(*)}$. Instead,  virtual states are obtained, see Table~\ref{tab:pole00} and Fig.~\ref{fig:specDDD1D100}. 

On the contrary, based on two prerequisites that 
\begin{itemize}
\item[1)]$X(3872)$ is a marginally bound state of $D^0\bar D^{*0}$ with binding energy $0\sim1$ MeV, and
\item[2)]$D_s\bar D_s$ can form a bound states with a binding energy of $2.4\sim 12.9$ MeV from the lattice result~\cite{Prelovsek:2020eiw} and the $\chi_{c0}(3930)$ mass determined by LHCb~\cite{Aaij:2020hon, Aaij:2020ypa},
\end{itemize}
Ref.~\cite{Meng:2020cbk} obtained $D_{s}^{(*)}\bar D_s^{(*)}$ bound systems with binding energies up to 80 MeV.

\begin{figure}[tb]
    \centering
    \includegraphics[width=\linewidth]{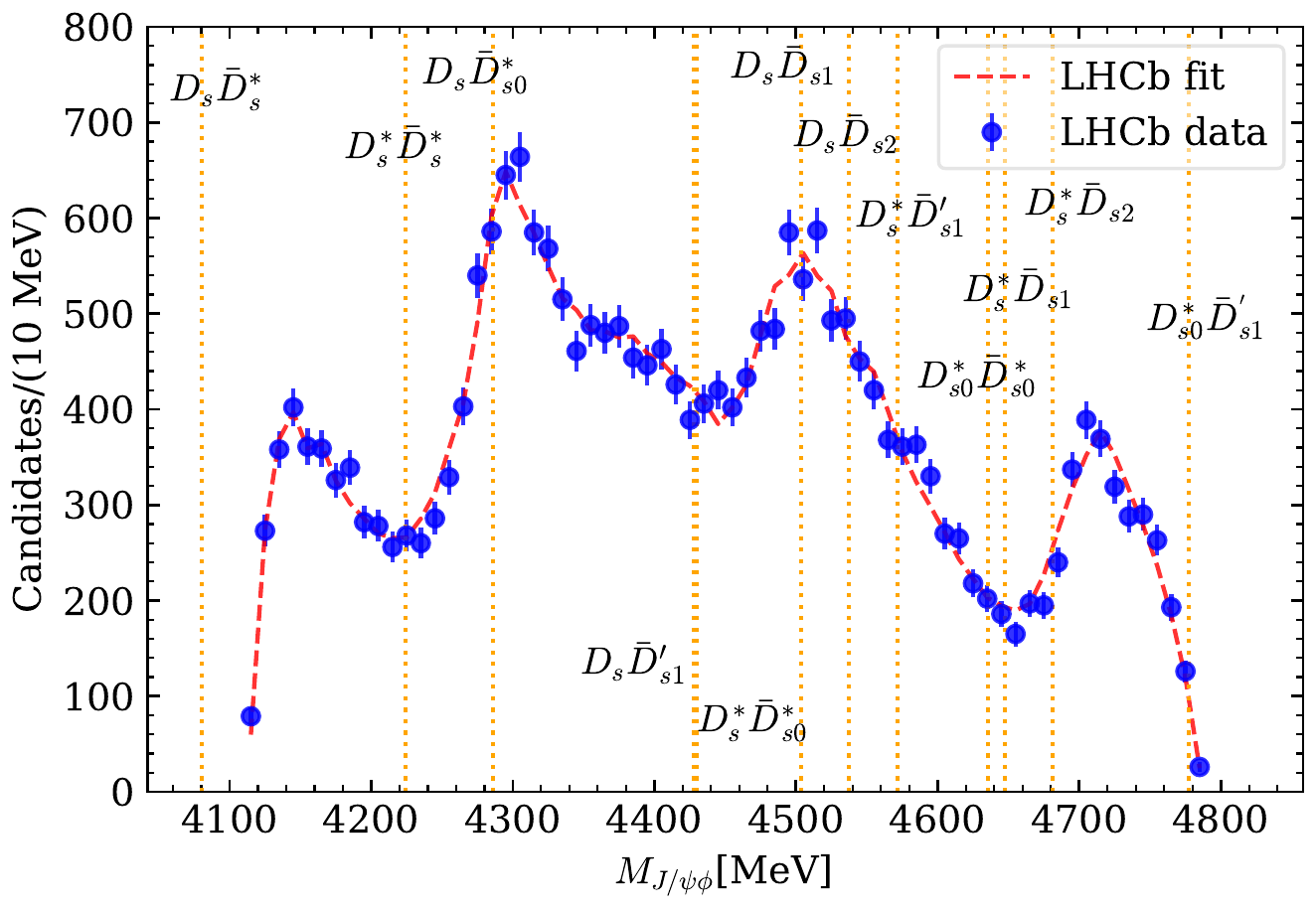}
    \caption{Thresholds of charm-strange meson pairs in the energy range relevant for the $B^+\to J/\psi \phi K^+$. Here, $D_{s0}^*$ denotes $D_{s0}^*(2317)$, $D_{s1}$ and $D_{s1}'$ denote $D_{s1}(2536)$ and $D_{s1}(2460)$, respectively, and $D_{s2}$ denotes $D_{s2}(2573)$.
    The data are taken from Ref.~\cite{Aaij:2016iza}.}
    \label{fig:jpsiphi}
\end{figure}
The $X(4140)$ first observed by the CDF Collaboration~\cite{Aaltonen:2009tz} was considered as a molecule of $D^*_s\bar D_s^*$ with $J^{PC}=0^{++}$ or $2^{++}$ in Refs.~\cite{Liu:2009ei,Branz:2009yt,Albuquerque:2009ak,Ding:2009vd,Zhang:2009st,Chen:2015fdn,Karliner:2016ith}, which is, however, disfavored by the results of LHCb~\cite{Aaij:2016iza,Aaij:2016nsc} where the $J^{PC}$ of $X(4140)$ were suggested to be $1^{++}$ (and thus the $X(4140)$ was named as $\chi_{c1}(4140)$ in the latest version of RPP). Actually in our calculation, it is not likely for the $D^*_s\bar D_s^*$ to form such a deeply bound state, noticing that the $X(4140)$ is about 80~MeV below the threshold of $D^*_s\bar D_s^*$. 
Instead, it is interesting to notice that just at the $D^*_s\bar D_s^*$ threshold there is evidence for a peak in the invariant mass distribution of $J/\psi \phi$, see Fig.~\ref{fig:jpsiphi}.\footnote{This structure around the  $D^*_s\bar D^*_s$ threshold drew the attention of Ref.~\cite{Wang:2017mrt} where two resonances, a narrow $X(4140)$ and a broad $X(4160)$, were introduced to fit the $J/\psi\phi$ invariant mass distribution from the threshold to
about 4250~MeV. There the broad $X(4160)$ was considered as a $D^*_s\bar D^*_s$ molecule.}
Following the analysis in Ref.~\cite{Dong:2020hxe}, if the interaction of the $D^*_s\bar D_s^*$ is attractive but not strong enough to form a bound state, a peak will appear just at the $D^*_s\bar D_s^*$ threshold in the invariant mass distribution of $J/\psi \phi$, and the peak is narrow if there is a nearby virtual state pole. 
A detailed study of this threshold structure can tell us whether the attraction between a pair of charm-strange mesons is strong enough to form a bound state or not.

The difference between the $J/\psi\phi$ and $D_s\bar D_s^*$ thresholds is merely 36~MeV. Thus, the shallow $D_s\bar D_s^*$ virtual state with $1^{++}$ could be responsible for the quick rise of the $J/\psi\phi$ invariant mass distribution just above threshold observed in the LHCb data~\cite{Aaij:2016iza}.

\subsubsection{$D^{(*)}\bar D_s^{(*)}$: $Z_{cs}$ as virtual states} 

No light vector can be exchanged here and the attractive interaction from vector charmonia exchange is crucial, similar to the isovector $D\bar D^{(*)}$ systems. A virtual state pole could exist below threshold. 
In particular, if the $Z_c(3900)$ exists as a virtual state, the same interaction would induce $D^{(*)}\bar D_s^{(*)}$ virtual states.

Recently, a near-threshold enhancement in the invariant mass distribution of $D_s^-D^{*0}+D_s^{*-}D^0$ was reported by the BESIII Collaboration~\cite{Ablikim:2020hsk} and an exotic state $Z_{cs}(3985)^-$ was claimed. This state has been widely investigated~\cite{Wang:2020dmv,Wan:2020oxt,Wang:2020kej,Meng:2020ihj,Yang:2020nrt,Chen:2020yvq,Du:2020vwb,Cao:2020cfx,Sun:2020hjw,Wang:2020rcx,Wang:2020htx,Wang:2020iqt,Azizi:2020zyq,Jin:2020yjn,Simonov:2020ozp,Sungu:2020zvk,Ikeno:2020csu,Xu:2020evn}, some of which regard it as a molecule of $D_s^-D^{*0}+D_s^{*-}D^0$ while some others object such an explanation. In Ref.~\cite{Yang:2020nrt}, it was found that a virtual or resonant pole together with a triangle singularity can well reproduce the line shape of the BESIII data, consistent with the analysis here.

\subsubsection{$D^{(*)}\bar D_{1,2}$: $Y(4260)$ and related states} 

It is possible for the isoscalar $D\bar D_1$ pair to form a bound state with a binding energy from a few MeV to dozens of MeV. Note that this system can have $J^{PC}=1^{-\pm}$ and the $1^{--}$ state is slightly more deeply bound than the $1^{-+}$ one which has exotic quantum numbers.

The $Y(4260)$ was discovered by the {\babar} Collaboration~\cite{Aubert:2005rm} with a mass of $(4259\pm8^{+2}_{-6})$ MeV and a width of 50 $\sim$ 90 MeV and later confirmed by other experiments~\cite{He:2006kg,Yuan:2007sj}. 
Now it is called $\psi(4230)$ due to a lower mass from the more precise BESIII data and a combined analysis in four channels, $e^{+} e^{-} \rightarrow \omega \chi_{c 0}$~\cite{Ablikim:2015uix}, $\pi^{+} \pi^{-} h_{c}$~\cite{BESIII:2016adj}, $\pi^{+} \pi^{-} J / \psi$~\cite{Ablikim:2016qzw} and $ D^{0} D^{*-} \pi^{+} + c.c.$~\cite{Ablikim:2018vxx}, yielding a mass of $(4219.6 \pm 3.3 \pm 5.1)$~MeV and a width of $(56.0 \pm 3.6 \pm 6.9)$~MeV~\cite{Gao:2017sqa}. This state is a good candidate of exotic states (see reviews, e.g., Refs.~\cite{Chen:2016qju,Guo:2017jvc,Brambilla:2019esw}). It was argued that the isoscalar $D\bar D_1$ plays important roles in the structure of $Y(4260)$ in, e.g., Refs.~\cite{Wang:2013cya,Qin:2016spb,Chen:2019mgp}. 
The binding energy of the isoscalar $D\bar D_1$ system with $J^{PC}=1^{-\pm}$ via the vector meson exchange was calculated by solving the Schr\"odinger equation in spatial space in Ref.~\cite{Dong:2019ofp}, and the results are consistent with this work except that the sign of $V_c$, which has a minor impact, is not correct there. Note that the mass of the isoscalar $D\bar D_1$ bound state  obtained here is larger than the nominally mass of the $\psi(4230)$, see Fig.~\ref{fig:specBB00}, but the mixing of the $D\bar D_1$ molecule with a $D$-wave vector charmonium~\cite{Lu:2017yhl} may solve this discrepancy.

From the results in Table~\ref{tab:pole00} and Fig.~\ref{fig:specBB00}, the isoscalar $D\bar D_1$ bound state has quite some partners, either of HQSS or of SU(3) flavor.
In particular, several of them have vector quantum numbers, including a $D^*\bar D_1$ bound state with a mass about 4.39 $\sim$ 4.43~GeV, a $D^*\bar D_2$ bound state with a mass about 4.43 $\sim$ 4.47~GeV, and three virtual states of  $D_s\bar D_{s1}$,  $D^*_s\bar D_{s1}$ and $D^*_s\bar D_{s2}$. 

The current status of the vector charmonium spectrum around 4.4~GeV is not clear, and the peak structures in exclusive and inclusive $R$-value measurements are different (for a compilation of the relevant data, see Ref.~\cite{Yuan:HAPOF}). Thus, it is unclear which structure(s) can be identified as the candidate(s) of the $D^*\bar D_{1(2)}$ bound states. Nevertheless, the $Y(4360)$, aka $\psi(4360)$, and $\psi(4415)$ have been suggested to correspond to the $D_1\bar D^*$ and $D_2\bar D^*$ states, respectively~\cite{Wang:2013kra,Ma:2014zva,Cleven:2015era,Hanhart:2019isz}.
A determination of the poles around 4.4~GeV would require a thorough analysis of the full data sets including these open-charm channels, and the first steps have been done in Refs.~\cite{Cleven:2013mka,Olschewsky:master}.

As for the virtual states with hidden-strangeness, they are expected to show up as narrow threshold cusps in final states like $J/\psi f_0(980)$ and $\psi(2S)f_0(980)$.
They could play an important role in generating the $Y(4660)$, aka $\psi(4660)$, peak observed in the $\psi(2S)f_0(980)\to \psi(2S)\pi^+\pi^-$ invariant mass distribution~\cite{Lees:2012pv,Wang:2014hta}.\footnote{The $Y(4660)$ was suggested to be a $\psi(2S)f_0(980)$ bound state in Ref.~\cite{Guo:2008zg} to explain why it was seen only the $\psi(2S)\pi^+\pi^-$ final state with the pion pair coming from the $f_0(980)$.}
Although it was proposed~\cite{Cotugno:2009ys,Guo:2010tk} that the $Y(4630)$ structure observed in the $\Lambda_c\bar \Lambda_c$ spectrum~\cite{Pakhlova:2008vn} could be the same state as the $Y(4660)$ one, the much more precise BESIII data~\cite{Ablikim:2017lct}, however, show a different behavior up to 4.6~GeV in the $\Lambda_c\bar \Lambda_c$ invariant mass distribution (see below).
Further complications come from the $1^{--}$ structures around 4.63~GeV reported in the $D_s\bar D_{s1}+c.c.$~\cite{Jia:2019gfe} and $D_s\bar D_{s2}+c.c.$~\cite{Jia:2020epr} distributions, the former of which has been proposed to be due to a molecular state from the $D_s^*\bar D_{s1}$-$D_s\bar D_{s1}$ interaction~\cite{He:2019csk}.
Suffice it to say that the situation of the $Y(4630)$ is not unambiguous. With more precise data that will be collected at BESIII and Belle-II, we suggest to search for line shape irregularities (either peaks or dips) at the $D_s^*\bar D_{s1}$ and $D_s^*\bar D_{s2}$ thresholds in open-charm-strangeness final states such as $D_s^{(*)}\bar D_s^{(*)}$ and $D_s\bar D_{s1(s2)}$.

There are also hints in data for positive $C$-parity $D_s\bar D_{s1}$ and $D_s^*\bar D_{s1}$ virtual states (see Table~\ref{tab:pole00c} and Fig.~\ref{fig:specBB00}), whose thresholds are at 4503 MeV and 4647~MeV, respectively. As can be seen from Fig.~\ref{fig:jpsiphi}, there is a peak around 4.51~GeV and a dip around 4.65~GeV in the $J/\psi\phi$ energy distribution measured by the LHCb Collaboration~\cite{Aaij:2016iza}, and the energy difference between the dip and peak approximately equals to the mass splitting between the $D_s^*$ and $D_s$. We also notice that the highest peak in the same data appears at the $D_{s0}^*(2317)\bar D_s$ threshold.\footnote{The coincidence of the peak position with the threshold and the highly asymmetric line shape suggests a $D_{s0}^*(2317)\bar D_s$ virtual state. Such systems will be studied in a future work.} 
All these channels, together with the $D_s\bar D_s^*$ and $D_s^*\bar D_s^*$ discussed in Section~\ref{sec:dsds}, need to be considered in a reliable analysis of the $B^+\to J/\psi\phi K^+$ data, which is however beyond the scope of this paper.

Notice that because the $D_{1(2)}$ and $D_{s1(s2)}$ have finite widths, the molecular states containing one of them can decay easily through the decays of $D_{1(2)}$ or $D_{s1(s2)}$. The structures at the thresholds of $D_{s}^{(*)}\bar D_{s1(s2)}$, for which virtual states are predicted, will get smeared by the widths of $D_{s1(s2)}$. Thus, the $D_{s}^{(*)}\bar D_{s2}$ threshold structures should be broader and smoother than the $D_{s}^{(*)}\bar D_{s1}$ ones since the width of the $D_{s2}$, $(16.9\pm0.7)$~MeV~\cite{Zyla:2020zbs}, is much larger than that of the $D_{s1}$, $(0.92\pm0.05)$~MeV~\cite{Zyla:2020zbs}.

\subsubsection{$\Lambda_c\bar \Lambda_c$: analysis of the BESIII data and more baryon-antibaryon bound states} 

From Table~\ref{tab:pole00} and Fig.~\ref{fig:specBB00}, in the spectrum of the isoscalar $1^{--}$ states, in addition to those made of a pair of charmed mesons, we predict more than 10 baryon-antibaryon molecules. The lowest one is the $\Lambda_c\bar \Lambda_c$ bound state, and the others are above 4.85~GeV. While those above 4.85~GeV are beyond the current reach of BESIII (there is a BESIII data-taking plan in the energy region above 4.6~GeV~\cite{Ablikim:2019hff}), there is strong evidence for the existence of a $\Lambda_c\bar \Lambda_c$ bound state in the BESIII data~\cite{Ablikim:2017lct}.

The $\Lambda_c\bar \Lambda_c$ system can form a bound state with a binding energy in the range from a few MeV to dozens of MeV, depending on the cutoff. Therefore, we predict that there is a pole below the $\Lambda_c\bar \Lambda_c$ threshold and the pole position can be extracted from the line shape of the  $\Lambda_c\bar \Lambda_c$ invariant mass distribution near threshold.

The cross section of $e^+e^-\to\Lambda_c\bar \Lambda_c$ was first measured using the initial state radiation by Belle~\cite{Pakhlova:2008vn} and a vector charmonium-like structure $Y(4630)$ was observed. The BESIII Collaboration measured such cross sections at four energy points just above threshold much more precisely~\cite{Ablikim:2017lct}. The energy dependence of the cross sections at these four points has a weird behavior: it is almost flat. 
This can be understood as the consequence of the Sommerfeld factor, which makes the distribution nonvanishing even exactly at threshold, and the existence of a near-threshold pole, which counteracts the increasing trend of the phase space multiplied by the Sommerfeld factor to result in an almost flat distribution.
Here we fit BESIII data to estimate where the pole is located.

The Sommerfeld factor~\cite{Sommerfeld:1931} accounting for the multi-photon exchange between the $\Lambda_c^+$ and $\bar \Lambda_c^-$ reads,
\begin{align}
S_0(E)=\frac{2\pi x}{1-e^{-2\pi x}},
\end{align}
where $x=\alpha\mu/k$ with $\alpha\approx1/137$, $\mu=m_{\Lambda_c}/2$, and $k$ is defined in Eq.~(\ref{eq:3-momentum}). The cross section of $e^+e^-\to\Lambda_c\bar \Lambda_c$ is now parameterized as
\begin{align}
\sigma(E)=N\cdot S_0(E)\cdot |f(E)|^2 \cdot\frac{\rho(E)}{E^2} ,
\end{align}
 with $N$ a normalization constant and $\rho(E)=k/(8\pi E)$ the phase space. Here $f(E)$ denotes the nonrelativistic scattering amplitude, and the $S$-wave one,
\begin{align}
f_0(E)=\left(\frac1{a_0}-i\sqrt{2\mu(E-2m_{\Lambda_c})}\right)^{-1},\label{eq:f}
\end{align}
is sufficient in the immediate vicinity of the threshold. Note that we take the scattering length $a_0$ complex to take into account the couplings between the $\Lambda_c\bar\Lambda_c$ and lower channels~\cite{Dong:2020hxe}. Finally, we have 3 parameters, Re$(1/a_0)$, Im$(1/a_0)$ and $N$, to fit four experimental data. 

\begin{figure}[tb]
    \centering
    \includegraphics[width=\linewidth]{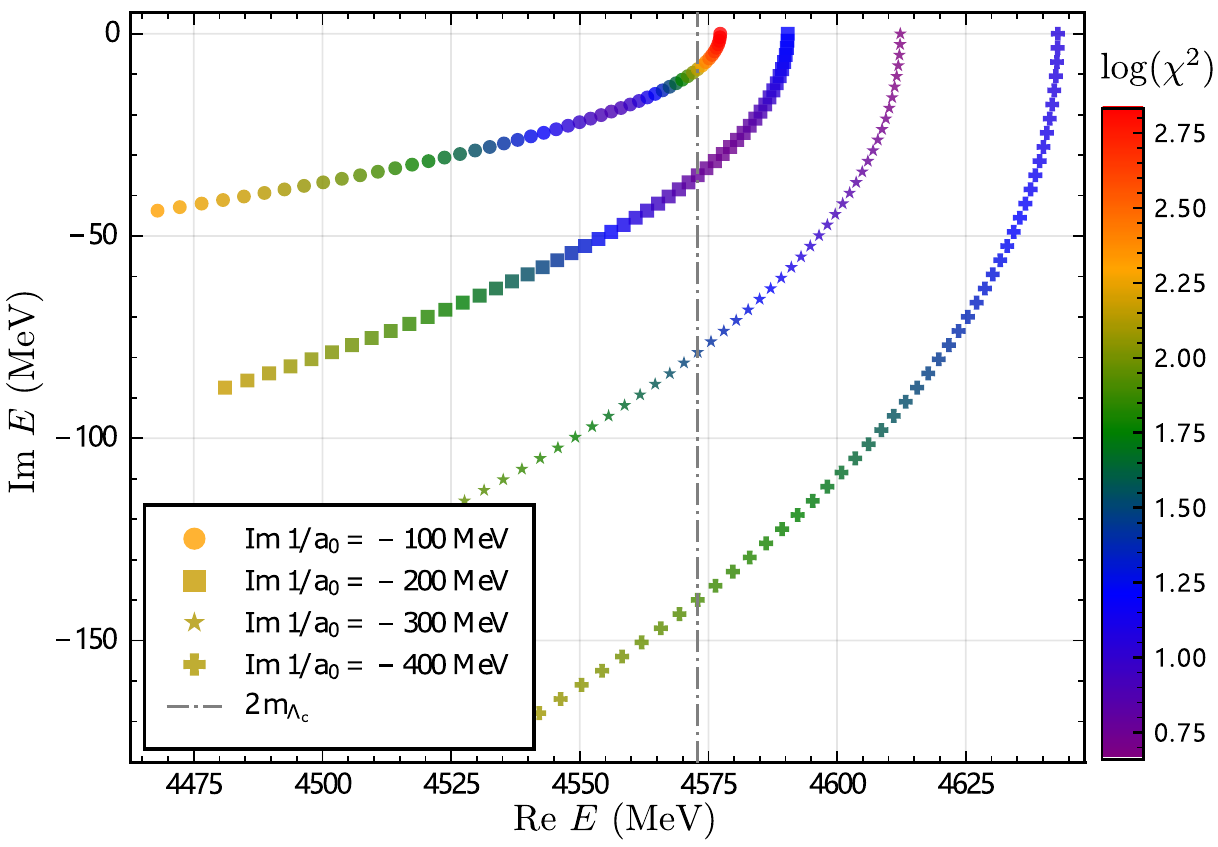}
    \includegraphics[width=\linewidth]{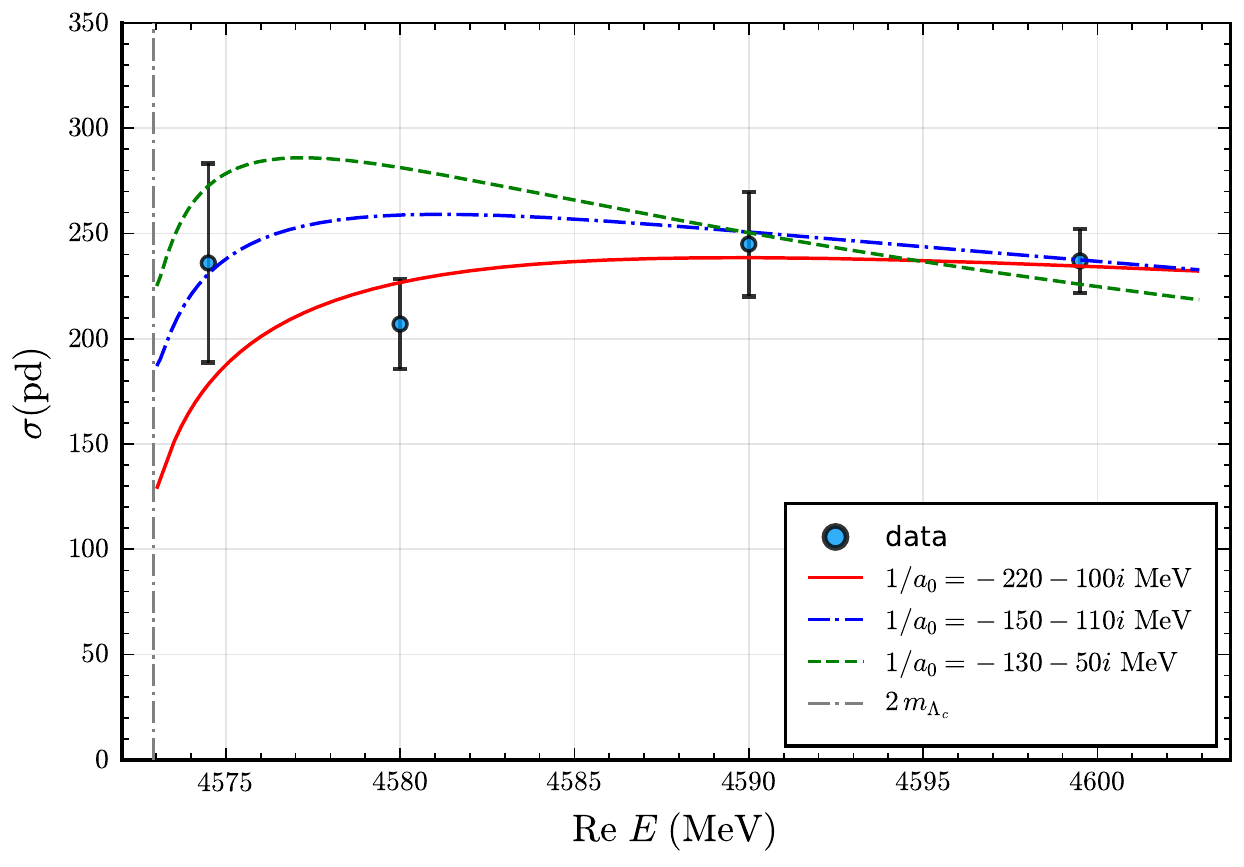}
    \caption{Top: pole positions of Eq.~\eqref{eq:f} on the first RS with different scattering length ($a_0$) values and the color represents the $\chi^2$ (in a logarithmic form for better illustration) of the fit to BESIII data~\cite{Ablikim:2017lct}. The pole on the second RS is at the same position if we change the sign of Re$(1/a_0)$, which does not change the fit. Bottom: examples of some fits, which yield poles below threshold at $4456-19i$ MeV (red), $4468-14i$ MeV (blue dash-dotted) and $4566-6i$ MeV (green dashed).}\label{fig:lambdac-ploe}
\end{figure}

The fitted results are shown in Fig.~\ref{fig:lambdac-ploe}. We can see that the best fit leads to a pole located close to the real axis but above threshold. A pole below threshold, as we predicted, is also possible, see the bottom one in Fig.~\ref{fig:lambdac-ploe}. These fits, though with larger $\chi^2$, are reasonable since we have only four points. The obtained $a_0$ values from these fits yield poles several MeV below threshold with an imaginary part of dozens of MeV. Such poles are located on the first RS corresponding to a bound state, which moves from the real axis onto the complex plane due to the coupling to lower channels. 
There is another pole located at the symmetric position on the second RS, corresponding to a virtual state. Actually, with the scattering length approximation in Eq.~\eqref{eq:f}, we cannot determine on which RS the pole is located since the poles on different RSs below threshold have the same behavior above threshold.
More data are needed to pin down the exact pole position corresponding to our predicted $\Lambda_c\bar \Lambda_c$ bound state, which should be different from the $Y(4630)$ or $Y(4660)$. 
In Ref.~\cite{Dai:2017fwx}, the BESIII~\cite{Ablikim:2017lct} and Belle~\cite{Pakhlova:2008vn} data are fitted together using an amplitude with a pole around 4.65~GeV. While the Belle data of the $Y(4630)$ peak can be well described, the much more precise BESIII data points in the near-threshold region cannot.
We conclude that the data from $\Lambda_c\bar \Lambda_c$ threshold up to 4.7~GeV should contain signals of at least two states: the $\Lambda_c\bar \Lambda_c$ molecule and another one with a mass around 4.65~GeV.

As for the isoscalar vector states above 4.85~GeV, the structures could be more easily identified from data than those around 4.3~GeV. This is because the charmonium states in that mass region should be very broad while these hadronic molecules are narrower due to the small binding energies, corresponding to large spatial extensions. We expect the $\Sigma_c\bar \Sigma_c$, $\Xi_c\bar \Xi_c$ and $\Sigma_c\bar \Sigma_c^*$ below 5~GeV to be seen in the forthcoming BESIII measurements, and the ones higher than 5~GeV can be searched for in future super tau-charm facilities~\cite{Barniakov:2019zhx,Peng:2020orp}.

There are isovector $1^{--}$ baryon-antibaryon molecular states above 4.7~GeV, see Table~\ref{tab:pole10} and Fig.~\ref{fig:BB10}. It is more difficult to observe these states than the isoscalar ones in $e^+e^-$ collisions since the main production mechanism of vector states should be driven by a vector $\bar c\gamma_\mu c$ current, which is an isoscalar, coupled to the virtual photon.
However, they could be produced together with a pion, and thus can be searched for in future super tau-charm facilities with center-of-mass energies above 5~GeV.

\subsubsection{$\bar D^{(*)} \Sigma_c^{(*)}$: $P_c$ states} 

The $\bar D^{(*)} \Sigma_c^{(*)}$ systems with isospin-$1/2$ are attractive, and a near-threshold pole can be found for each combination. This pole is a virtual state or (mostly) a bound state depending on the cutoff, see Table~\ref{tab:pole50} and Fig.~\ref{fig:specDB50}.

Such systems have drawn lots of attention~\cite{Wu:2010jy,Wu:2010vk,Wu:2010rv,Wang:2011rga,Yang:2011wz,Wu:2012md,Xiao:2013yca,Karliner:2015ina} especially after the pentaquark states, $P_c(4450)$ and $P_c(4380)$, were observed by LHCb~\cite{Aaij:2015tga}. In the updated measurement~\cite{Aaij:2019vzc}, the $P_c(4450)$ signal splits into two narrower peaks, $P_c(4440)$ and $P_c(4457)$. There is no clear evidence for the previous broad $P_c(4380)$, and meanwhile a new narrow resonance $P_c(4312)$ shows up. Several models have been applied by tremendous works to understand the structures of these states, and the $\bar D^{(*)} \Sigma_c^{(*)}$ molecular explanation stands out as it can explain the three states simultaneously, see e.g. Refs.~\cite{Liu:2019tjn,Xiao:2019aya,Du:2019pij}. Particularly in Ref.~\cite{Du:2019pij}, the LHCb data are described quite well by the interaction constructed with heavy quark spin symmetry and actually four $P_c$ states, instead of three, show up, corresponding to $\bar D\Sigma_c$, $\bar D\Sigma_c^*$ and $\bar D^*\Sigma_c$ molecules.
A hint of a narrow $P_c(4380)$ was reported in the analysis of Ref.~\cite{Du:2019pij}. The rest three $P_c$ states related to $\bar D^*\Sigma_c^*$ predicted there have no signals up to now. 

In the vector meson saturation model considered here, the two contact terms constructed considering only HQSS~\cite{Liu:2018zzu,Liu:2019tjn,Sakai:2019qph,Du:2019pij}, corresponding to the total angular momentum of the light degrees of freedom to be $1/2$ and $3/2$, are the same, similar to the $H\bar H$ interaction discussed in Section~\ref{sec:DD}. 
As a result, 7 $\bar D^{(*)}\Sigma_c^{(*)}$ molecular states~\cite{Xiao:2013yca} with similar binding energies are obtained, and the two $\bar D^*\Sigma_c$ states with different total spins, corresponding to the $P_c(4440)$ and $P_c(4457)$, degenerate. 
The degeneracy will be lifted by considering the exchange of pion and other mesons and keeping the momentum-dependent terms of the light vector exchange.

\subsubsection{$\bar D^{(*)}\Xi_c^{(\prime)}$: $P_{cs}$ and related states} 

It is natural for the isoscalar $\bar D^*\Xi_c$ to form bound states if the above $P_c$ states are considered as the isospin-$1/2$ $\bar D^{(*)} \Sigma_c^{(*)}$ molecules since the interactions from the light vector exchange are the same in these two cases, see Table~\ref{tab:potentials}. Note that such states have been predicted by various works~\cite{Hofmann:2005sw,Chen:2016ryt,Anisovich:2015zqa,Wang:2015wsa,Feijoo:2015kts,Lu:2016roh,Xiao:2019gjd,Chen:2015sxa,Wang:2019nvm,Zhang:2020cdi}.

Recently, Ref.~\cite{Aaij:2020gdg} reported an exotic state named $P_{cs}(4459)$ in the invariant mass distribution of $J/\psi\Lambda$ in $\Xi_b^-\to J/\psi K^-\Lambda$. Even though the significance is only 3.1$\sigma$, several works~\cite{Liu:2020hcv,Chen:2020kco,Wang:2020eep,Peng:2020hql,Chen:2020uif} have explored the possibility of $P_{cs}(4459)$ being a molecule of $\bar D^* \Xi_c$, and the finding here supports such an explanation that the structure could be caused by two isoscalar $\bar D^*\Xi_c$ molecules.

Furthermore, Ref.~\cite{Wang:2020bjt} moved forward to the double strangeness systems and claimed that $\bar D_s^* \Xi_c'$ and $\bar D_s^*\Xi_c^*$  may form bound states with $J^P=3/2^-$ and $5/2^-$, respectively. The $\phi$ exchange for such systems yields repulsive interaction at leading order, see Table~\ref{tab:potentials}, and the bound states obtained in Ref.~\cite{Wang:2020bjt} result from other contributions, including the exchange of pseudoscalar and scalar mesons, the subleading momentum dependence from the $\phi$ exchange and coupled-channel effects. 

\section{Summary and discussion}
\label{sec:5}

The whole spectrum of hadronic molecules of a pair of charmed and anticharmed  hadrons, considering all the $S$-wave singly-charmed mesons and baryons as well as the $s_\ell=3/2$ $P$-wave charmed mesons, is systematically obtained using $S$-wave constant contact potentials  saturated by the exchange of vector mesons. 
The coupling of charmed heavy hadrons and light mesons are constructed by implementing HQSS, chiral symmetry and SU(3) flavor symmetry. 

The spectrum predicted here should be regarded as the leading approximation of the spectrum for heavy-antiheavy molecular states, 
and gives only a general overall feature of the heavy-antiheavy hadronic molecular spectrum. Specific systems may differ from the predictions here due to the limitations of our treatment.
We considered neither the effects of coupled channels, nor the spin-dependent interactions, which arises from momentum-dependent terms that are of higher order in the very near-threshold region, nor the contribution from the exchange of pseudoscalar and scalar mesons, nor the mixing with charmonia. Nevertheless, the spectrum shows a different pattern than that considering only the one-pion exchange (see, e.g., Ref.~\cite{Karliner:2015ina}), which does not allow the molecular states in systems such as $D\bar D$ and $\Sigma_c\bar D$, where the one-pion exchange is forbidden without coupled channels, to exist.  

In total 229 hidden-charm hadronic molecules (bound or virtual) are predicted, many of which deserve attentions:

\begin{itemize}

    \item[1)] The pole positions of the isoscalar $D\bar D^*$ with positive and negative $C$-parity are consistent with the molecular explanation of $X(3872)$ and $\tilde X(3872)$, respectively. There is a shallow bound state in the isoscalar $D\bar D$ system, consistent with the recent lattice QCD result~\cite{Prelovsek:2020eiw}.
    
    \item[2)] The spectrum of the $\bar D^{(*)} \Sigma_c^{(*)}$ systems is consistent with the molecular explanations of famous $P_c$ states: the $P_c(4312)$ as an isospin-$1/2$ $\bar D \Sigma_c$ molecule, and $P_c(4440)$ and $P_c(4457)$ as isospin-$1/2$ $\bar D^* \Sigma_c$ molecules. With the resonance saturation from the vector mesons, the two $\bar D^* \Sigma_c$ molecules are degenerated. In addition, there is an isospin-$1/2$ $\bar D\Sigma_c^*$ molecule, consistent with the narrow $P_c(4380)$ advocated in Ref.~\cite{Du:2019pij}, and three isospin-$1/2$ $\bar D^*\Sigma_c^*$ molecules, consistent with the results in the literature.
   
    \item[3)] There are two isoscalar $\bar D^*\Xi_c$ molecules, which may be related to the recently announced $P_{cs}(4459)$. In addition, more negative-parity isoscalar $P_{cs}$-type molecules are predicted: one in $\bar D\Xi_c$, one in $\bar D\Xi_c'$, one in $\bar D \Xi_c^*$, two in $\bar D^*\Xi_c'$, and three in $\bar D^*\Xi_c^*$.

    \item[4)] Instead of associating the $X(4140)$  with a $D_s^*\bar D_s^*$ molecule like some other works did, our results prefer the $D_s^*\bar D_s^*$ to form a virtual state. The peak in the invariant mass distribution of $J/\psi \phi$ measured by LHCb just at the $D_s^*\bar D_s^*$ threshold is consistent with this scenario, according to the discussion in Ref.~\cite{Dong:2020hxe}.
    
    \item[5)] The isoscalar $D^{(*)}\bar D_{1(2)}$ can form negative-parity bound states with both positive and negative $C$ parities. The $D\bar D_1$ bound state is the lowest one in this family, and the $1^{--}$ one is consistent with the sizeable $D\bar D_1$ molecular component in the $\psi(4230)$.
    
    \item[6)] $\Lambda_c\bar\Lambda_c$ bound states with $J^{PC}=0^{-+}$ and $1^{--}$ are predicted. 
    The vector one should be responsible to the almost flat line shape of the $e^+e^-\to\Lambda_c\bar\Lambda_c$ cross section in the near-threshold region observed by BESIII~\cite{Ablikim:2017lct}. 
    
    \item[7)] Light vector meson exchanges either vanish due to the cancellation between $\rho$ and $\omega$ or are not allowed in the isovector $D^{(*)}\bar D^{(*)}$ systems and $D^{(*)}\bar D_s^{(*)}$ systems. 
    However, the vector charmonia exchanges may play an important role as pointed out in Ref.~\cite{Aceti:2014uea}, and the $Z_c(3900,4020)$ and $Z_{cs}(3985)$ could well be the $D^{(*)}\bar D^*$ and $D^*\bar D_s$ -- $D\bar D_s^*$ virtual states.
  
\end{itemize}

When the light vector meson exchange is allowed, the results reported here are generally consistent with the results from a more complete treatment of the one-boson exchange model (e.g., by solving the Schr\"odinger equation). 
For example, the binding energy of the isoscalar  $D\bar D$~\cite{Zhang:2006ix} and $D\bar D_1$~\cite{Dong:2019ofp} bound states from the $\rho$ and $\omega$ exchanges fit the spectrum well;
a similar pattern of molecular states related to the $X(3872)$ was obtained in Ref.~\cite{Liu:2008tn} and the light vector exchange was found necessary to bind $D\bar D^*$ together;
the $\bar D^*\Sigma_c$ bound states corresponding to the $P_c(4440)$ and $P_c(4457)$ were obtained via one boson exchange in Ref.~\cite{Liu:2019zvb}, and the degeneracy of the two states with $J=1/2$ and $3/2$ is lifted by the pion exchange and higher order contributions from the $\rho$ and $\omega$ exchange.
We should also notice that there can be systems whose contact terms receive important contributions from the scalar-meson exchanges.

We expect that there should be structures in the near-threshold region for all the heavy-antiheavy hadron pairs that have attractive interactions at threshold. The structure can be either exactly at threshold, if the attraction is not strong enough to form a bound state, or below threshold, if a bound state is formed. Moreover, the structures are not necessarily peaks, and they can be dips in invariant mass distributions, depending on the pertinent production mechanism as discussed in our recent work~\cite{Dong:2020hxe}. 

When the predicted states have ordinary quantum numbers as those of charmonia, the molecular states must mix with charmonia, and the mixing can have an important impact on the spectrum. Yet, in the energy region higher than 4.8~GeV, where plenty of states are predicted as shown in Figs.~\ref{fig:specBB00} and \ref{fig:specDDD1D100}, normal charmonia should be very broad due to the huge phase space while the molecular states should be relatively narrow due to the large distance between the consistent hadrons. Thus, narrow structures to be discovered in this energy region should be mainly due to the molecular structures, being either bound or virtual states.

Among the 229 structures predicted here, only a minority is in the energy region that has been studied in detail. The largest data sets from the current experiments have the following energy restrictions: direct production of the $1^{--}$ sector in $e^+e^-$ collisions goes up to 4.6~GeV at BESIII; the hidden-charm $XYZ$ states produced through the weak process $b\to c\bar c s$ in $B\to K$ decays should be below 4.8~GeV; the hidden-charm $P_c$ pentaquarks produced in $\Lambda_b\to K$ decays should be below 5.1~GeV.
To find more states in the predicted spectrum, we need to have both data in these processes with higher statistics and data at other experiments such as the prompt production at hadron colliders, PANDA, electron-ion collisions and $e^+e^-$ collisions above 5~GeV at super tau-charm facilities.

The potentials in the bottom sector are the same as those in the charm sector, if using the nonrelativistic field normalization, due to the HQFS, and we expect the same number of molecular states in the analogous systems therein. Because of the much heavier reduced masses of hidden-bottom systems, the virtual states in the charm sector will move closer to the thresholds or even become bound states in the bottom sector, and the bound states in the charm sector will be more deeply bound in the bottom sector. There may even be excited states for some deeply bound systems.
For these deeply bound systems, the constant contact term approximation considered here will not be sufficient.
However, due to the large masses, such states are more difficult to be produced than those in the charm sector.

\begin{acknowledgments}
We would like to thank Chang-Zheng Yuan for useful discussions, and thank Fu-Lai Wang for a communication regarding Ref.~\cite{Wang:2020bjt}.
This work is supported in part by the Chinese Academy of Sciences (CAS) under Grant No.~XDB34030000 and No.~QYZDB-SSW-SYS013, by the National Natural Science Foundation of China (NSFC) under Grant No.~11835015, No.~12047503 and No.~11961141012, by the NSFC and the Deutsche Forschungsgemeinschaft (DFG, German Research Foundation) through the funds provided to the Sino-German Collaborative Research Center TRR110 ``Symmetries and the Emergence of Structure in QCD'' (NSFC Grant No. 12070131001, DFG Project-ID 196253076), and by the CAS Center for Excellence in Particle Physics (CCEPP).
\end{acknowledgments}

\begin{appendix}

\section{Vertex factors for direct processes\label{app:vertex}}
The vertex factors in Eq.~(\ref{eq:Amp}) for different particles are calculated in the following:
\begin{itemize}
    \item $D$
    \begin{align}
        g_1&=-\sqrt2 \beta g_V m_D (P_a^{(Q)} V_{ab} P^{(Q)T}_b),\\
        g_2&=\sqrt2 \beta g_V m_D (P_a^{(\bar Q)} V_{ab} P^{(\bar Q)T}_b).
    \end{align}
    \item $D^*$
        \begin{align}
        g_1&=\sqrt2 \beta g_V m_{D^*}\epsilon\cdot\epsilon^*(P_a^{*(Q)} V_{ab} P^{*(Q)T}_b)\notag\\
&\approx-\sqrt2 \beta g_V m_{D^*}(P_a^{*(Q)} V_{ab} P^{*(Q)T}_b),  \\
        g_2&=-\sqrt2 \beta g_V m_{D^*}\epsilon\cdot\epsilon^*(P_a^{*(\bar Q)} V_{ab} P^{*(\bar Q)T}_b)\notag\\
        &\approx\sqrt2 \beta g_V m_{D^*}(P_a^{*(\bar Q)} V_{ab} P^{*(\bar Q)T}_b). 
    \end{align}
    \item $D_1$
    \begin{align}
        g_1&=-\sqrt2 \beta_2 g_V m_{D_1} \epsilon\cdot\epsilon^*  (P^{(Q)}_{1a} V_{ab} P^{(Q)T}_{1b}),\notag\\
&\approx \sqrt2 \beta_2 g_V m_{D_1} (P^{(Q)}_{1a} V_{ab} P^{(Q)T}_{1b}),\\
        g_2&=\sqrt2 \beta_2 g_V m_{D_1} \epsilon\cdot\epsilon^*  (P^{(\bar Q)}_{1a} V_{ab} P^{(\bar Q)T}_{1b}),\notag\\
&\approx -\sqrt2 \beta_2 g_V m_{D_1} (P^{(\bar Q)}_{1a} V_{ab} P^{(\bar Q)T}_{1b}).
    \end{align}
    \item $D_2^*$
        \begin{align}
        g_1&=\sqrt2 \beta_2 g_V m_{D_2^*}\epsilon^{\mu\nu}\epsilon_{\mu\nu}^*(P^{*(Q)}_{2a} V_{ab} P^{*(Q)T}_{2b})\notag\\
&\approx\sqrt2 \beta_2 g_V m_{D_2^*}(P^{*(Q)}_{2a} V_{ab} P^{*(Q)T}_{2b}),  \\
        g_2&=-\sqrt2 \beta g_Vm_{D_2^*} \epsilon^{\mu\nu}\epsilon_{\mu\nu}^*(P^{*(\bar Q)}_{2a} V_{ab} P^{*(\bar Q)T}_{2b})\notag\\
        &\approx\sqrt2 \beta g_V m_{D_2^*}(P^{*(\bar Q)}_{2a} V_{ab} P^{*(\bar Q)T}_{2b}). 
    \end{align}
    \item $B_{\bar 3}$
        \begin{align}
        g_1&=\frac1{\sqrt2} \beta_B g_V \bar u(k_1)u(p_1){\rm tr}\lrb{( B_{\bar3}^{(Q)T} V B_{\bar3}^{(Q)}}\notag\\
&\approx  \sqrt2 \beta_B g_V m_{B_{\bar3}} {\rm tr}\lrb{(B_{\bar3}^{(Q)T} V B_{\bar3}^{(Q)}}, \\
        g_2&=-\frac1{\sqrt2} \beta_B g_V \bar u(k_2)u(p_2){\rm tr}\lrb{ B_{3}^{(\bar Q)T} V^T B_{3}^{(\bar Q)}}\notag\\
        &\approx-\sqrt2 \beta_B g_V m_{B_{\bar3}}{\rm tr}\lrb{ B_{3}^{(\bar Q)T} V^T B_{3}^{(\bar Q)}} .
    \end{align}
    \item $B_6$
        \begin{align}
        g_1&=-\frac13\frac{\beta_S g_V}{\sqrt2}\bar u(k_1)\gamma^5(\gamma^\mu+v^\mu)^2\gamma^5u(p_1)\notag\\
        &\quad\times{\rm tr}\lrb{B_{6}^{(Q)T}VB^{(Q)}_{6}}\notag\\
&\approx-\sqrt2m_{B_6}\beta_S g_V{\rm tr}\lrb{B_{6}^{(Q)T}VB^{(Q)}_{6}}, \\
        g_2&=\frac13\frac{\beta_S g_V}{\sqrt2}\bar u(k_2)\gamma^5(\gamma^\mu+v^\mu)^2\gamma^5u(p_2)\notag\\
         &\quad\times{\rm tr}\lrb{B_{6}^{(\bar Q)T}V^TB^{(\bar Q)}_{6}}\notag\\
        &\approx\sqrt2m_{B_6}\beta_S g_V{\rm tr}\lrb{B_{6}^{(\bar Q)T}V^TB^{(\bar Q)}_{6}}.
    \end{align}
    \item $B_6^*$
        \begin{align}
        g_1&=\frac{\beta_S g_V}{\sqrt2} \bar u_{\mu}^*(k_1)u^{*\mu}(p_1){\rm tr}\lrb{B_{6}^{*(Q)T}VB_{6}^{*(Q)}}\notag\\
&\approx-\sqrt2m_{B_6^*}\beta_S g_V {\rm tr}\lrb{B_{6}^{*(Q)T}VB_{6}^{*(Q)}},\\
        g_2&=-\frac{\beta_S g_V}{\sqrt2} \bar u_{\mu}^*(k_2)u^{*\mu}(p_2){\rm tr}\lrb{B_{6}^{*(\bar Q)T}V^TB_{6}^{*(\bar Q)}}\notag\\
        &\approx\sqrt2m_{B_6^*}\beta_S g_V {\rm tr}\lrb{B_{6}^{*(\bar Q)T}V^TB_{6}^{*(\bar Q)}}.
    \end{align}
\end{itemize}
In the above deductions we have used $\epsilon\cdot\epsilon^*=-1$, $\epsilon^{\mu\nu}\cdot\epsilon_{\mu\nu}^*=1$, $\bar u(k_1)u(p_1)=2m$ and $\bar u_{6\mu}^{*} u_6^{*\mu}=-2m$ at threshold. Note that the factors such as $P_a^{(Q)} V_{ab} P^{(Q)T}_b$, $ {\rm tr}\lrb{ B_{\bar3}^{(Q)T} V B_{\bar3}^{(Q)}}$ in the above expressions contain only the SU(3) flavor information and the properties of the corresponding fields have already been extracted.

\section{List of the potential factor $F$}\label{app:poten}

The details of interactions between all combinations of heavy-antiheavy hadron pairs are listed in Tables~\ref{tab:potentials} and \ref{tab:potentials1}. 

\begin{table*}[tbhp]
\caption{The group theory factor $F$, defined in Eq.~\eqref{eq:potential}, for the interaction of charmed-anticharmed hadron pairs with only the light vector-meson exchanges. Here both charmed hadrons are the $S$-wave ground states. $I$ is the isospin and $S$ is the strangeness. Positive $F$ means attractive. For the systems with $F=0$, the sub-leading exchanges of  vector-charmonia also lead to an attractive potential at threshold. } \label{tab:potentials}
\centering
\begin{ruledtabular}
\begin{tabular}{l|cccc}
System &  $(I,S)$ & Thresholds (MeV) & Exchanged particles & $F$\\
\hline
$D^{(*)}\bar D^{(*)}$& (0,0)&$(3734,3876,4017)$ &$\rho,\omega$ & $\frac32,\frac12$\\
& (1,0)& &$\rho,\omega$ & $-\frac12,\frac12$\\
$D_s^{(*)}\bar D^{(*)}$& $(\frac12,1)$&$(3836,3977,3979,4121)$&$-$ & $0$\\
$D^{(*)}_s\bar D^{(*)}_s $& (0,0)&$(3937,4081,4224)$&$\phi$ & $1$\\
\hline
$\bar D^{(*)}\Lambda_c$& $(\frac12,0)$ &$(4154,4295)$&$\omega$ & $-1$\\
$\bar D_s^{(*)}\Lambda_c$& $(0,-1)$ &$(4255,4399)$&$-$ & $0$\\
$\bar D^{(*)}\Xi_c$& $(1,-1)$ &$(4337,4478)$ &$\rho,\omega$ & $-\frac12,-\frac12$\\
& $(0,-1)$& &$\rho,\omega$ & $\frac32,-\frac12$\\
$\bar D_s^{(*)}\Xi_c$& $(\frac12,-2)$& $(4438,4582)$&$\phi$ & $-1$\\
\hline
$\bar D^{(*)}\Sigma_c^{(*)}$& $(\frac32,0)$&$(4321,4385,4462,4527)$ &$\rho,\omega$ & $-1,-1$\\
& $(\frac12,0)$& &$\rho,\omega$ & $2,-1$\\
$\bar D_s^{(*)}\Sigma_c^{(*)}$& $(1,-1)$&$(4422,4486,4566,4630)$ &$-$ & $0$\\
$\bar D^{(*)}\Xi_c^{'(*)}$& $(1,-1)$&$(4446,4513,4587,4655)$ &$\rho,\omega$ & $-\frac12,-\frac12$\\
& $(0,-1)$& &$\rho,\omega$ & $\frac32,-\frac12$\\
$\bar D_s^{(*)}\Xi_c^{'(*)}$& $(\frac12,-2)$& $(4547,4614,4691,4758)$&$\phi$ & $-1$\\
$\bar D^{(*)}\Omega_c^{(*)}$& $(\frac12,-2)$&$(4562,4633,4704,4774)$ &$-$ & $0$\\
$\bar D_s^{(*)}\Omega_c^{(*)}$& $(0,-3)$&$(4664,4734,4807,4878)$ &$\phi$ & $-2$\\

\hline
$ \Lambda_c\bar\Lambda_c$& $(0,0)$&$(4573)$ &$\omega$ & $2$\\
$\Lambda_c\bar \Xi_c$& $(\frac12,1)$&$(4756)$ &$\omega$ & $1$\\
$\Xi_c\bar \Xi_c$& $(1,0)$&$(4939)$ &$\rho,\omega,\phi$ & $-\frac12,\frac12,1$\\
& $(0,0)$& &$\rho,\omega,\phi$ & $\frac32,\frac12,1$\\
\hline
$\Lambda_c\bar\Sigma_c^{(*)}$& $(1,0)$&$(4740,4805)$ &$\omega$ & $2$\\

$\Lambda_c\bar\Xi_c^{'(*)}$&$(\frac12,1)$&$(4865,4932)$ &$\omega$ & $1$\\

$\Lambda_c\bar\Omega_c^{(*)}$ &$(0,2)$&$(4982,5052)$ &$-$ & $0$\\

$\Xi_c \bar\Sigma_c^{(*)}$  &$(\frac32,-1)$&$(4923,4988)$ &$\rho,\omega$ & $-1,1$\\

&$(\frac12,-1)$& &$\rho,\omega$ & $2,1$\\
$\Xi_c \bar\Xi_c^{'(*)}$ &$(1,0)$&$(5048,5115)$ &$\rho,\omega,\phi$ & $-\frac12,\frac12,1$\\

& $(0,0)$& &$\rho,\omega,\phi$ & $\frac32,\frac12,1$\\
$\Xi_c \bar\Omega_c^{(*)}$ &$(\frac12,1)$&$(5165,5235)$ &$\phi$ & $2$\\
\hline
$\Sigma_c^{(*)}\bar\Sigma_c^{(*)}$  & $(2,0)$&$(4907,4972,5036)$ &$\rho,\omega$ & $-2,2$\\

 & $(1,0)$& &$\rho,\omega$ & $2,2$\\

 & $(0,0)$& &$\rho,\omega$ & $4,2$\\

$\Sigma_c^{(*)}\bar\Xi^{'(*)}_c$  &$(\frac32,1)$&$(5032,5097,5100,5164)$ &$\rho,\omega$ & $-1,1$\\

 & $(\frac12,1)$& &$\rho,\omega$ & $2,1$\\

$\Sigma_c^{(*)}\bar\Omega^{(*)}_c$  &$(0,2)$& $(5149,5213,5219,5284)$&$-$ & $0$\\

$\Xi_c^{'(*)} \bar\Xi_c^{'(*)}$&$(1,0)$&$(5158,5225,5292)$ &$\rho,\omega,\phi$ & $-\frac12,\frac12,1$\\

 &$(0,0)$& &$\rho,\omega,\phi$ & $\frac32,\frac12,1$\\

$\Xi^{'(*)}_c \bar\Omega_c^{(*)}$&$(\frac12,1)$&$(5272,5341,5345,5412)$ &$\phi$ & $2$\\

$\Omega_c ^{(*)}\bar\Omega_c^{(*)}$ &$(0,0)$&$(5390,5461,5532)$ &$\phi$ & $4$
\end{tabular}
\end{ruledtabular}
\end{table*}

\begin{table*}
\caption{The group theory factor $F$, defined in Eq.~\eqref{eq:potential}, for the interaction of charmed-anticharmed hadron pairs with only the light vector-meson exchanges. Here one of the charmed hadrons is an $s_\ell=3/2$ charmed meson. See the caption of Table~\ref{tab:potentials}. } \label{tab:potentials1}
\centering
\begin{ruledtabular}
\begin{tabular}{l|cccc}
System &  $(I,S)$ & Thresholds (MeV) & Exchanged particles & $F$ \\
\hline
$D^{(*)}\bar D_{1,2}$& (0,0)&$(4289,4330,4431,4472)$ &$\rho,\omega$ & $\frac32,\frac12$\\
& (1,0)& &$\rho,\omega$ & $-\frac12,\frac12$\\
$D^{(*)}\bar D_{s1,s2}$& $(\frac12,-1)$&$(4390,4431,4534,4575)$&$-$ & $0$\\
$D_s^{(*)}\bar D_{1,2}$& $(\frac12,1)$&$(4402,4436,4544,4578)$&$-$ & $0$\\
$D^{(*)}_s\bar D_{s1,s2} $& (0,0)&$(4503,4537,4647,4681)$&$\phi$ & $1$\\
\hline
$D_{1,2}\bar D_{1,2}$& (0,0)&$(4844,4885,4926)$ &$\rho,\omega$ & $\frac32,\frac12$\\
& (1,0)& &$\rho,\omega$ & $-\frac12,\frac12$\\
$D_{s1,s2}\bar D_{1,2}$& $(\frac12,1)$&$(4957,4991,4998,5032)$&$-$ & $0$\\
$D_{s1,s2}\bar D_{s1,s2} $& (0,0)&$(5070,5104,5138)$&$\phi$ & $1$\\
\hline
$\Lambda_c\bar D_{1,2}$& $(\frac12,0)$ &$(4708,4750)$&$\omega$ & $-1$\\
$\Lambda_c\bar D_{s1,s2}$& $(0,-1)$ &$(4822,4856)$&$-$ & $0$\\
$\Xi_c\bar D_{1,2}$& $(1,-1)$ &$(4891,4932)$ &$\rho,\omega$ & $-\frac12,-\frac12$\\
& $(0,-1)$& &$\rho,\omega$ & $\frac32,-\frac12$\\
$\Xi_c\bar D_{s1,s2}$& $(\frac12,-2)$& $(5005,5039)$&$\phi$ & $-1$\\
\hline
$\Sigma_c^{(*)}\bar D_{1,2}$& $(\frac32,0)$&$(4876,4917,4940,4981)$ &$\rho,\omega$ & $-1,-1$\\
& $(\frac12,0)$& &$\rho,\omega$ & $2,-1$\\
$\Sigma_c^{(*)}\bar D_{s1,s2}$& $(1,-1)$&$(4989,5023,5053,5087)$ &$-$ & $0$\\
$\Xi_c^{'(*)}\bar D_{1,2}$& $(1,-1)$&$(5001,5042,5068,5109)$ &$\rho,\omega$ & $-\frac12,-\frac12$\\
& $(0,-1)$& &$\rho,\omega$ & $\frac32,-\frac12$\\
$\Xi_c^{'(*)}\bar D_{s1,s2}$& $(\frac12,-2)$& $(5114,5148,5181,5215)$&$\phi$ & $-1$\\
$\Omega_c^{(*)}\bar D_{1,2}$& $(\frac12,-2)$&$(5117,5158,5188,5229)$ &$-$ & $0$\\
$\Omega_c^{(*)}\bar D_{s1,s2}$& $(0,-3)$&$(5230,5264,5301,5335)$ &$\phi$ & $-2$\\
\end{tabular}
\end{ruledtabular}
\end{table*}

\section{Amplitude calculation for cross processes\label{app:cross}}

In the following we show the deduction of Eq.~(\ref{eq:Vc}).
%and the factor $F_c$ is collected in Table~\ref{tab:Fc}.

\begin{itemize}
\item $D\bar D_1$ and $D_s\bar D_{s1}$
\begin{align}
V&=i \lra{i\frac{-2\zeta_1g_V}{\sqrt3}}\lra{i\frac{2\zeta_1g_V}{\sqrt3}}m_{D}m_{D_1}\notag\\
&\times\epsilon_1^{*\mu}\frac{-i(g_{\mu\nu}-q_\mu q_\nu/m_{\rm ex}^2)}{q^2-m_{\rm ex}^2+i\epsilon}\epsilon_2^\nu F\notag\\
&\approx FF_c \zeta_1^2g_V^2\frac{m_{D}m_{D_1}}{m_{\rm ex}^2-\Delta{m}^2}
\end{align}
with $F_c=4/3$.

\item $D^*\bar D_1$ and $D_s^*\bar D_{s1}$

\begin{align}
V&=i\lra{i\frac{i\zeta_1g_V}{\sqrt3}}\lra{i\frac{-i\zeta_1g_V}{\sqrt3}m_{D^*}m_{D_1}}\epsilon_{\alpha\beta\gamma\delta}\epsilon_{\alpha_1\beta_1\gamma_1\delta_1}\notag\\
&\times \epsilon_1^\beta\epsilon_2^{*\alpha}v^{\gamma}\frac{-i(g^{\delta\delta_1}-q^\delta q^{\delta_1}/m_{\rm ex}^2)}{q^2-m_{\rm ex}^2+i\epsilon}\epsilon_3^{\alpha_1}\epsilon_4^{*\beta_1}v^{\gamma_1}F\notag\\
&\approx\frac13 \zeta_1^2g_V^2 \frac{m_{D^*}m_{D_1}}{m_{\rm ex}^2-\Delta{m}^2}(\bm \epsilon_1\times\bm \epsilon_2^*)\cdot(\bm \epsilon_3\times\bm \epsilon_4^*)F\notag\\
&=-\frac13 F\zeta_1^2g_V^2 \frac{m_{D^*}m_{D_1}}{m_{\rm ex}^2-\Delta{m}^2}\bm S_1\cdot \bm S_2\notag\\
&= FF_c\zeta_1^2g_V^2 \frac{m_{D^*}m_{D_1}}{m_{\rm ex}^2-\Delta{m}^2},
\end{align}
where $\bm S_i$ is the spin-1 operator. Explicitly, $\bm S_1\cdot \bm S_2=-2,-1$ and $1$ for total spin $J=0,1$ and $2$, respectively and in turn we obtain $F_c=2/3,1/3$ and $-1/3$.

\item  $D^*\bar D_2$ and $D_s^*\bar D_{s2}$

\begin{align}
V&=i\lra{i\sqrt2\zeta_1g_V}\lra{i\sqrt2\zeta_1g_V}m_{D^*}m_{D_2}\notag\\
&\times\epsilon_{1\mu}\epsilon_2^{*\mu\nu}\frac{-i(g_{\nu\alpha}-q_\nu q_\alpha)/m_{\rm ex}^2}{q^2-m_{\rm ex}^2+i\epsilon}\epsilon_3^{\alpha\beta}\epsilon_{4\beta}F\notag\\
&\approx {2\zeta_1^2g_V^2} \frac{m_{D^*}m_{D_2}}{m_{\rm ex}^2-\Delta{m}^2}\epsilon_{1\mu}\epsilon_2^{*\mu\nu}\epsilon_{3\nu}^{\beta}\epsilon_{4\beta}F\notag\\
&=FF_c\zeta_1^2g_V^2 \frac{m_{D^*}m_{D_2}}{m_{\rm ex}^2-\Delta{m}^2},
\end{align}
where $F_c=-1/3,1$ and $-2$ for total spin $J=1,2$ and $3$, respectively.
\end{itemize}

\iffalse
\begin{table}[t]
    \caption{The group theory factor $F_c$ defined in Eq.~(\ref{eq:Vc}). $J$ refers to the total spin.} \label{tab:Fc}
    \centering
    \begin{ruledtabular}
      \begin{tabular}{c|cccc}
        $J$&  0 & 1 & 2 & 3  \\\hline
       $D_{(s)}D_{(s)1}$ & 0 & $4/3$ & 0 & 0 \\
       $D^*_{(s)}D_{(s)1}$ & $2/3$ & $1/3$ & $-1/3$& 0  \\
       $D^*_{(s)}D_{(s)2}$ & 0 & $-1/3$ & $1$ & $-2$\\
      \end{tabular}
    \end{ruledtabular}
\end{table}
\fi

\end{appendix}

\medskip

\bibliography{ref}

\end{document}